\documentstyle[rmp,aps]{revtex}

\begin{document} 

\def\ltorder{\mathrel{\raise.3ex\hbox{$<$}\mkern-14mu
             \lower0.6ex\hbox{$\sim$}}}
\def\gtorder{\mathrel{\raise.3ex\hbox{$>$}\mkern-14mu
             \lower0.6ex\hbox{$\sim$}}}
\def\lapr{~\hbox{$\ltorder$}~}
\def\gapr{~\hbox{$\gtorder$}~}
\def\etal {{\it et al. }}
\def\etalv{{\it et al., }}
\def\etals{{\it et al.}'s }
\def\hy{H~{\sc i} }
\def\h2{H~{\sc ii}}
\def\is{interstellar }
\def\mf{magnetic field}
\def\mfs{magnetic fields}
\def\crs{cosmic rays}
\def\c_r{cosmic-ray }
\def\sni{Type~I supernovae}
\def\sn2{Type~II supernovae}

\title{The Interstellar Environment of our Galaxy}
\author{Katia M. Ferri\`ere}
\address{Observatoire Midi-Pyr\'en\'ees, 14 av. Ed. Belin,
           31400 Toulouse, France\footnote{ferriere@ast.obs-mip.fr}}
\maketitle

\begin{abstract}
We review the current knowledge and understanding of the interstellar
medium of our galaxy. We first present each of the three basic constituents 
-- ordinary matter, \crs, and magnetic fields --
of the interstellar medium, laying emphasis on their physical and chemical
properties inferred from a broad range of observations.
We then position the different interstellar constituents,
both with respect to each other and with respect to stars,
within the general galactic ecosystem. 
\end{abstract}

\tableofcontents

\section{Introduction}
\label{s1}

The stars of our galaxy 
-- traditionally referred to as ``the Galaxy" with a capital G 
to distinguish it from the countless other galaxies --
are embedded in an extremely tenuous medium, 
the so-called ``interstellar medium" (ISM), which contains ordinary
matter, relativistic charged particles known as \crs, and \mfs.
These three basic constituents have comparable pressures 
and are intimately coupled together by electromagnetic forces.
Through this coupling, \crs \ and \mfs \ influence both the dynamics
of the ordinary matter and its spatial distribution at all scales,
providing, in particular, an efficient support against the gravitational
force.
Conversely, the weight of the ordinary matter confines \mfs \ and,
hence, \crs \ to the Galaxy, while its turbulent motions can be held
responsible for the amplification of \mfs \ and for the acceleration of \crs.

The ISM encloses but a small fraction of the total mass of the Galaxy.
Moreover, it does not shine in the sky as visibly and brightly as stars do.
Yet, it plays a vital role in many of the physical and
chemical processes taking place in the Galactic ecosystem.

The most important aspect of Galactic ecology is probably the cycle 
of matter from the ISM to stars and back to the ISM.
In the first step of this cycle, new stars form out of a reservoir 
of \is material. This material, far from being uniformly spread
throughout \is space, displays dramatic density and temperature
contrasts, such that only the densest, coldest molecular regions
can offer an environment favorable to star formation.
In these privileged sites, pockets of \is gas, losing part of their 
magnetic support, tend to become gravitationally unstable and collapse
into new stars.

Once locked in the interior of stars, the Galactic matter goes through
a succession of thermonuclear reactions, which enrich it
in heavy elements. A fraction of this matter eventually returns to the ISM,
be it in a continuous manner via powerful stellar winds,
or in an instantaneous manner upon supernova explosions
(violent stellar outbursts resulting from a thermonuclear instability 
or from the sudden gravitational collapse of the core of some stars 
at the end of their lifetime).
In both cases, the injection of stellar mass into the ISM is accompanied
by a strong release of energy, which, in addition to generating
turbulent motions in the ISM, contributes to maintaining its highly 
heterogeneous structure and may, under certain circumstances,
give birth to new molecular regions prone to star formation.
This last step closes the loop of the partly self-induced ISM-star
cycle.

Thus, the ISM is not merely a passive substrate within which stars evolve; 
it constitutes their direct partner in the Galactic ecosystem,
continually exchanging matter and energy with them, and controlling
many of their properties.
It is the spatial distribution ot the \is material together with
its thermal and chemical characteristics  that determines the locations
where new stars form as well as their mass and luminosity spectra.
These, in turn, govern the overall structure, the optical appearance,
and the large-scale dynamics of the Galaxy.
Hence understanding the present-day properties of our Galaxy and being
able to predict its long-term evolution requires a good knowledge of 
the dynamics, energetics, and chemistry of the ISM.

The purpose of the present review is precisely to describe the current
status of this knowledge. Major advances have been made over the last
few years, thanks, in large part, to steady improvements 
in instrumentation, in observation and analysis techniques, 
and in computer power. A number of recent large-scale surveys 
in different wavelength bands have allowed astronomers to obtain
complementary images with unprecedented spatial coverage, resolution,
and sensitivity, often supplemented by valuable spectral information.
These surveys have also stimulated numerous theoretical studies, 
aimed at interpreting the observed phenomena within the context 
of a well-understood model and using them to place observational
constraints on the various processes at work.

We will start with a brief historical overview, emphasizing the main
developments that paved the way to the modern view of our Galaxy
(Section~\ref{s2}).
We will then present the three basic constituents of the ISM,
namely, the ordinary matter (Section~\ref{s4}), the \crs \ and
the \mfs \ (Section~\ref{s5}).
Finally, we will discuss the interplay between these three constituents
and their relations with stars (Section~\ref{s6}).

\section{Historical Background}
\label{s2}

\subsection{Overall Picture of the Galaxy}
\label{s2A}

To a terrestrial observer, the Galaxy appears (by starry nights only) 
as a faint band of diffuse light stretching all the way around the sky; 
this is why it is also known as the Milky Way galaxy or simply the Milky Way.
This denomination goes back to the ancient Greek civilization
and, in particular, to Claude Ptolemy (90 -- 168), who provided
one of the very first descriptions of the Milky Way, qualifying it as
``a zone as white as milk".
However, the true nature of the Milky Way was not established until 1610,
when Galileo Galilei (1564 -- 1642), examining it for the first time
through his telescope, discovered that it was actually composed of
innumerable dim stars.

It became clear in the course of the eighteenth century that 
we see the Milky Way as a narrow band encircling us because it has
the shape of a flattened disk, deep into which we are embedded.
At the end of the century, William Herschel (1738 -- 1822) undertook
a systematic study of the distribution of stars across the sky.
Since he relied on a couple of faulty assumptions 
-- all stars have approximately the same intrinsic brightness 
and \is space is completely transparent to starlight -- he erroneously 
concluded that the Galaxy is about five times more extended in its plane 
than in the perpendicular direction and that the Sun is located near 
the Galactic center.
These conclusions were corroborated a good century later by Jacobus
Kapteyn (1851 -- 1922), based on the far more abundant stellar data
available at his time. Going one step further, Kapteyn also estimated 
the spatial distribution of stars within the Galaxy together with
its overall size; he thus obtained exponential scale lengths 
at half maximum of $\sim 1.2$~kpc in the radial direction and 
$\sim 0.22$~kpc in the vertical direction.\footnote{
Because of the Earth's annual revolution about the Sun, a nearby star
seems to trace out an ellipse in the plane of the sky with respect to
the very distant background stars. The parallax is defined as the angle
under which the semi-major axis of this apparent ellipse is seen from Earth, 
and the parsec (pc) is, by definition, the length unit equal to the distance 
at which a star has a parallax of one second of arc.
This length unit is the most commonly used by astronomers.
Nonetheless, in the context of individual galaxies, it often proves more
convenient to employ the kiloparsec (kpc). For future reference,
1~kpc = 1000~pc = 3260~light-years = $3.09 \times 10^{16}$~km.}

A totally different picture of the Galaxy emerged during World War~I,
from Harlow Shapley's (1885 -- 1972) observational work on globular
clusters (compact, nearly spherical groupings of $10^5$ to $10^7$ stars).
He noticed that, unlike ordinary stars, globular clusters do not spread
uniformly along the Milky Way, but instead concentrate in the direction
of the Sagittarius constellation. He further found that they have 
a roughly spherical distribution, the center of which, he argued, 
should approximately coincide with the center of the Galaxy. 
This is how his investigation led him to the radical conclusion that the Sun
lies very far from the Galactic center, at a distance of about 15~kpc.

Strong support in favor of Shapley's picture came in the mid 1920's
from the kinematical studies of Bertil Lindblad (1895 -- 1965) 
and Jan Oort (1900 -- 1992),
which convincingly showed that the observed relative velocities of stars
and globular clusters with respect to the Sun were readily understood 
in the framework of a differentially rotating Galactic model with the Sun 
placed at the radial distance predicted by Shapley, whereas they were hard
to reconcile with the total amount of mass inferred from Kapteyn's model. 
Almost three decades later, radio-astronomical measurements of 
the spatial distribution of \is neutral hydrogen delivered
the definitive proof that Shapley was correct about the off-center 
position of the Sun in the Galaxy, while demonstrating that he had 
overestimated its Galactocentric radius by almost a factor of 2.

We now know that our Galaxy comprises a thin disk with radius
$\sim 25 - 30$~kpc and effective thickness $\sim 400 - 600$~pc,
plus a spherical system itself composed of 
a bulge with radius $\sim 2 - 3$~kpc and a halo extending out to 
more than 30~kpc from the center (Binney and Merrifield, 1998, p.~606). 
The Sun resides in the Galactic disk, approximately 15~pc above the midplane 
(Cohen, 1995; Magnani \etalv 1996) and 8.5~kpc away from the center 
(Kerr and Lynden-Bell, 1986).

The stars belonging to the disk rotate around the Galactic center in nearly
circular orbits. Their angular rotation rate is a decreasing function 
of their radial distance. At the Sun's orbit, the rotation velocity
is $\simeq 220$~km~s$^{-1}$ (Kerr and Lynden-Bell, 1986), corresponding 
to a rotation period of about 240 million years.
Disk stars also have a velocity dispersion $\sim 10 - 40$~km~s$^{-1}$ 
(Mihalas and Binney, 1981, p.~423), which causes them to execute small
oscillations about a perfectly circular orbit, both in the Galactic plane
(epicycles) and in the vertical direction.
In contrast, the stars present in the bulge and in the halo rotate slowly 
and often have very eccentric trajectories.

Radio-astronomical observations of \is neutral hydrogen indicate that 
the Milky Way
possesses a spiral structure, similar to that seen optically in numerous 
external galaxies. These ``spiral galaxies" typically
exhibit two spiral arms unwinding either directly from the central bulge 
or from both ends of a bar crossing the bulge diametrically.
The exact spiral shape of our own Galaxy is difficult to determine 
from within; the best-to-date radio data point to a structure characterized 
by a bulge of intermediate size and a moderate winding of the arms 
(type Sbc in Hubble's classification; Binney and Merrifield, 1998, p.~171), 
while recent infrared (IR) images of the Galactic center region clearly
display the distinctive signature of a bar (Blitz and Spergel, 1991;
Dwek \etalv 1995).
Our position with respect to the spiral pattern
can be derived from local optical measurements, which give a quite
accurate outline of the three closest arms; they locate the Sun 
between the inner Sagittarius arm and the outer Perseus arm,
near the inner edge of the local Orion-Cygnus arm (Mihalas and Binney,
1981, p.~248).\footnote{
In fact, the local Orion-Cygnus arm is probably a short spur
rather than a major spiral arm like Sagittarius and Perseus
(Blitz \etalv 1983).}

\subsection{The Interstellar Medium}
\label{s2B}

The Milky Way system is not only made of stars; it also contains
significant amounts of tenuous matter, inhomogeneously spread out
throughout \is space. The \is matter, which exists
in the form of gas (atoms, molecules, ions, and electrons) and dust
(tiny solid particles), manifests itself primarily through obscuration,
reddening, and polarization of starlight, through the formation
of absorption lines in stellar spectra, and through various emission
mechanisms (both over a continuum and at specific wavelengths).
It is, incidentally, the presence of obscuring \is material
that gave Herschel and Kapteyn the false impression that the spatial
density of stars falls off in all directions away from us and, 
thus, brought them to misplace the Sun near the center of the Galaxy.
Shapley did not encounter the same problem with globular clusters, 
both because they are intrinsically much brighter and easier to recognize
than individual stars and because most of them lie outside the thin layer 
of obscuring material.

Herschel, back in the late eighteenth century, had already noticed that
some regions in the sky, particularly along the Milky Way, seemed 
devoid of stars. The first long-exposure photographs of the Milky Way 
taken by Edward Barnard (1857 -- 1923) in the early days of astronomical
photography revealed many more dark zones with a variety of shapes and sizes.
It was soon realized that these apparent holes in the stellar distribution
were due to the presence, along the line of sight, of discrete ``clouds" 
of \is matter hiding the stars situated behind them.
More specifically, it is the \is dust contained in these 
``dark clouds" that either absorbs or scatters the background starlight, 
the combination of these two processes being commonly called
\is obscuration or extinction.

Astronomers also suspected the existence of less conspicuous 
``diffuse clouds", especially after Hartmann's (1904) discovery 
of stationary absorption lines of once ionized calcium (Ca~{\sc ii}) 
in the spectrum of the spectroscopic binary $\delta$~Orionis.
Like in any spectroscopic binary system, the spectral lines created by 
the two companion stars orbiting around each other undergo a periodically 
varying Doppler shift, resulting from 
the back-and-forth motion of the stars along the line of sight.
Hence the stationary Ca~{\sc ii} lines could not arise from 
$\delta$~Orionis itself, but instead must have an \is origin.
In addition, their single-peaked shape together with their narrow width
strongly suggested that they were produced in a single cloud of ``cold"
($T < 1000$~K) \is gas somewhere between $\delta$~Orionis and the Earth.
The subsequent detection of absorption lines with multiple narrow peaks
in the spectrum of other stars provided further evidence for the existence 
of cold \is gas clumped into distinct clouds,
the line multiplicity being naturally attributed to the presence 
of several intervening clouds with different line-of-sight velocities
(e.g., Beals, 1936; Adams, 1949).

Shortly after the existence of \is clouds had been firmly established, 
Trumpler (1930) demonstrated that the space between the clouds was, 
in turn, filled with a widespread \is material. His argument 
rested on a comprehensive analysis of the properties of open clusters 
(rather loose, irregular groupings of $10^2$ to $10^3$ stars, confined
to the Galactic disk and, therefore, also known as Galactic clusters). 
He first estimated the distance to each of the observed clusters
by calculating the ratio of the apparent brightness of the most luminous
stars in the cluster to their intrinsic brightness, itself deduced from
their spectral type, and by assuming that \is space is transparent 
to starlight. He then multiplied the measured angular diameter of 
the cluster by its estimated distance, in order to evaluate its true size.
Proceeding in this manner, he found a systematic tendency for the more
distant clusters to be larger, regardless of the considered direction. 
Since this tendency could not be considered real -- otherwise the Sun would 
have a special position within the Galaxy -- Trumpler was led to conclude
that the light from remote clusters is gradually dimmed,
as it propagates through \is space, by the obscuring action 
of a pervasive \is material. Here, too, the precise agent
causing this general obscuration is \is dust.

The obscuration process due to \is dust is more effective
at shorter wavelengths, which are closer to the typical grain sizes
(see Section~\ref{s4F}), so that blue light is more severely dimmed 
than red light. In consequence, the light emitted by a far-away star
appears to us redder than it actually is. This reddening effect can 
be measured by comparing the observed apparent color of the star to 
the theoretical color corresponding to its spectral type.
By means of such measurements, 
Trumpler (1930) was able to show that the reddening of stars of a given
spectral type increases with their distance from us, thereby bringing 
the conclusive proof that \is space is indeed pervaded by 
an obscuring, dust-bearing, \is material.

Another manifestation of \is dust, equally linked to its obscuration
properties, is the linear polarization of starlight. 
The polarization effect, uncovered about two decades after the reddening
effect, is easily understood if \is dust particles are elongated and
partially aligned by a large-scale \mf \ (Davis and Greenstein, 1951).
Interpreted in this manner, the observed polarization of starlight
furnished the first solid piece of evidence that the ISM is threaded
by coherent \mfs.

It took a few more years to realize that the ISM is filled with \crs \ too.
Although the existence of \crs \ outside the Earth's atmosphere had been
known since the balloon experiment conducted by Hess (1919), the Galactic
origin of the most energetic of them and their widespread distribution
throughout the Milky Way were not recognized until the observed Galactic
radio emission was correctly identified with synchrotron radiation by 
\c_r electrons gyrating about \mf \ lines (Ginzburg and Syrovatskii, 1965).

\section{Interstellar Matter}
\label{s4}

\subsection{General Properties}
\label{s4A}

The \is matter accounts for $\sim 10 - 15~\%$ of the total mass 
of the Galactic disk. It tends to concentrate near the Galactic plane and 
along the spiral arms, while being very inhomogeneously distributed at small
scales. Roughly half the \is mass is confined to discrete
clouds occupying only $\sim 1 - 2~\%$ of the \is volume.
These \is clouds can be divided into three types: the dark clouds, 
which are essentially made of very cold ($T \sim 10 - 20$~K) molecular gas
and block off the light from background stars, 
the diffuse clouds, which consist of cold ($T \sim 100$~K) atomic gas
and are almost transparent to the background starlight, except at a number
of specific wavelengths where they give rise to absorption lines,
and the translucent clouds, which contain molecular and atomic gases
and have intermediate visual extinctions.
The rest of the \is matter, spread out between the clouds, 
exist in three different forms: warm (mostly neutral) atomic,
warm ionized, and hot ionized, where warm refers to a temperature
$\sim 10^4$~K and hot to a temperature $\sim 10^6$~K 
(see Table~\ref{t1}, below).

By terrestrial standards, the \is matter is exceedingly tenuous:
in the vicinity of the Sun, its density varies from 
$\sim 1.5 \times 10^{-26}$~g~cm$^{-3}$ in the hot medium
to $\sim 2 \times 10^{-20} - 2 \times 10^{-18}$~g~cm$^{-3}$ in the densest 
molecular regions, with an average of about $2.7 \times 10^{-24}$~g~cm$^{-3}$
(see next subsections).
This mass density, which corresponds to approximately one hydrogen atom
per cubic centimeter, is over twenty orders of magnitude smaller than 
in the Earth's lower atmosphere.

The chemical composition of \is matter is close to the ``cosmic
composition" inferred from abundance measurements in the Sun, in other
disk stars, and in meteorites, namely, 90.8~\% by number [70.4~\% by mass] 
of hydrogen, 9.1~\% [28.1~\%] of helium, and 0.12~\% [1.5~\%] of heavier 
elements, customarily termed ``metals" in the astrophysical community 
(Spitzer, 1978, p.~4).
However, observations of \is absorption lines in the spectra of hot stars
indicate that a significant fraction of these heavier elements is often
missing or ``depleted" from the gaseous phase of the ISM, being, in all
likelihood, locked up in solid dust grains. Since the first systematic 
studies of \is elemental abundances along different sight lines
(Morton \etalv 1973; Rogerson \etalv 1973), depletion factors have been
known to vary appreciably across the sky, presumably due to the wide 
fluctuations in environmental physical conditions. As a general rule,
depletions tend to be more severe in regions with higher density and 
lower temperature (Jenkins, 1987; Van Steenberg and Shull, 1988);
they also seem to depend weakly on the ionization degree, insofar as
they are somewhat less in the warm ionized medium than in the warm neutral 
medium (Howk and Savage, 1999).
On the average, the most common ``metals", C, N, and O, are only depleted
by factors $\sim 1.2 - 3$, whereas refractory elements like Mg, Si, and Fe
are depleted by factors $\sim 10 - 100$ (Savage and Sembach, 1996).
Altogether, about $0.5 - 1~\%$ of the \is matter by mass is in the form 
of dust rather than gas.

In the following subsections, we focus on the \is gas and successively 
describe the five different forms under which it can be found: molecular, 
cold atomic, warm atomic, warm ionized, and hot ionized.
The last subsection is devoted to a description of the \is dust.

\subsection{Molecular Gas}
\label{s4B}

The first \is molecules (CH, CH$^+$, and CN) were discovered in
the late 1930's, through the optical absorption lines they produce in
stellar spectra. However, it was not until 1970, when ultraviolet (UV)
astronomy from above the Earth's atmosphere had just opened a new window 
on the Universe, that the most abundant \is molecule, H$_2$, 
was for the first time detected in the far-UV spectrum of a hot star 
(Carruthers, 1970).
The next most abundant molecule, CO, was identified in a UV stellar
spectrum the following year (Smith and Stecher, 1971).

These discoveries were succeeded in 1972 by the launch of a UV spectrometer 
on the {\it Copernicus} satellite, which prompted many observational 
studies on the \is molecular gas (see Spitzer and Jenkins, 1975, 
for a preliminary review of the {\it Copernicus} results).
The {\it Copernicus} survey of H$_2$ absorption by Savage \etal (1977)
provided the H$_2$ column density (number of H$_2$ molecules in a cylinder
of unit cross section along the line of sight) between the Earth and
109 nearby hot stars and led to first estimates of the space-averaged
(i.e., smoothed-out) density and temperature of the molecular gas near the Sun.

A wealth of additional information on the spatial distribution and
physical properties of the molecular gas is expected from 
the {\it Far Ultraviolet Spectroscopic Explorer (FUSE)} satellite,
which was launched in 1999 and will measure key absorption lines in 
the far-UV spectrum of hundreds of Galactic and extragalactic sources, 
with a much higher sensitivity than {\it Copernicus} (Moos \etalv 2000);
early release results on \is H$_2$ have been reported by Shull \etal (2000)
and Snow \etal (2000).

Observations of optical and UV absorption lines, crucial as they are
in our grasp of \is molecules, do not allow astronomers 
to probe the interior of dense molecular clouds, for the bright sources
necessary to make absorption measurements are obscured by the \is
dust present in the very regions to be probed.
In order to explore the structure and large-scale distribution of 
the molecular gas, one may take advantage of the fact that radio waves
are not subject to \is extinction and appeal to radio spectroscopy.

The H$_2$ molecule itself is not directly observable at radio wavelengths: 
because it possesses no permanent electric dipole moment and has a very
small moment of inertia, all its permitted transitions lie outside the
radio domain (Field \etalv 1966).
The CO molecule, for its part, has a $J = 1 \to 0$ rotational transition
at a radio wavelength of 2.6~mm; the corresponding emission line, which
was first observed a few months before the detection of CO in UV absorption 
(Wilson \etalv 1970), has become the primary tracer of molecular \is gas 
(e.g., Scoville and Sanders, 1987).
The technique employed to deduce the molecular spatial distribution 
at Galactic scales from the profile of the CO 2.6-mm emission line 
relies on the Galactic rotation curve; a detailed description of the method 
is given in the Appendix.

The first large-scale surveys of CO 2.6-mm emission, carried out by
Scoville and Solomon (1975) and by Burton \etal (1975),
covered only a thin band along the Galactic equator over longitude
intervals (as defined in Fig.~\ref{galaxy}) accessible from the northern 
terrestrial hemisphere. Nevertheless, they already
showed that most of the molecular gas resides in a well-defined ring
extending radially between 3.5~kpc and 7~kpc from the Galactic center,
and they unveiled a strong molecular concentration in the region 
interior to 0.4~kpc.\footnote{
\label{note}
These early surveys assumed $R_\odot = 10$~kpc for the Galactocentric
radius of the Sun, whereas the last IAU recommended value is 8.5~kpc
(Kerr and Lynden-Bell, 1986).
In consequence, we scaled down the lengths inferred from these surveys
by a factor of 0.85. Likewise, throughout this paper, all parameters 
taken from observational studies of the spatial distribution of
\is constituents will systematically be rescaled to 
$R_\odot = 8.5$~kpc. For reference, distances scale as $R_\odot$,
surface densities as $R_\odot^0$, volume densities as $R_\odot^{-1}$,
and masses as $R_\odot^2$.}

Subsequent, more extensive CO surveys allowed a finer description 
of the molecular gas radial distribution, and added information on 
its azimuthal and vertical distributions as well as on its small-scale 
structure.
Dame \etal (1987) assembled data from 5 large and 11 more restricted
surveys to construct a synoptic picture of the whole Milky Way.
Although they made no attempt to convert the dependence on line-of-sight 
velocity (which may be considered an observable,
being directly related to the measurable Doppler shift) 
into a dependence on heliocentric distance 
(by means of the Galactic rotation curve; see Appendix),
they were able to bring to light the spiral pattern of CO emission.
Indeed, they found that CO concentrations in the longitude--velocity
plane tend to follow the strips corresponding to the spiral arms
observed in the 21-cm emission of neutral \is hydrogen
(see Section~\ref{s4C}).
Let us mention that, in the continuation of Dame \etals (1987) work,
Dame \etal (2001) produced a new composite CO survey of the entire
Galactic disk with about $2 - 4$ times better angular resolution
and up to 10 times higher sensitivity per unit solid angle.

The precise horizontal distribution of the molecular material
in the first Galactic quadrant (quadrant~I in Fig.~\ref{galaxy})
was investigated by Clemens \etal (1988).
They contrived a means to overcome the near--far distance ambiguity 
for emission inside the solar circle (see Appendix), which enabled them
to draw a detailed face-on map of \is CO. This map is dominated 
by the molecular ring peaking at a Galactic radius $R \simeq 4.5$~kpc 
and, to a lesser extent,
by two discrete features closely associated with 21-cm spiral arms.
Interarm regions, though, are not devoid of molecular gas: their inferred 
H$_2$ space-averaged density is, on average over the first quadrant,
only a factor $\sim 3.6$ lower than in the arms.

The situation is apparently different in the outer Galaxy 
($R > R_\odot$). CO surveys of relatively extended portions of the sky
have provided detailed images in which molecular concentrations 
clearly line up along the 21-cm spiral arms. 
However, the molecular surface density contrast ratios between spiral arms 
and interarm regions are much greater than in the inner Galaxy, 
with a mean value $\sim 13:1$ 
both in the longitude range $270^\circ - 300^\circ$ (Grabelsky \etalv 1987)
and in the longitude range $102^\circ - 142^\circ$ (Heyer, 1999).
Such large density contrasts imply that, outside the solar circle,
the bulk of the molecular gas belongs to the spiral arms.
  
Bronfman \etal (1988) combined two separate CO surveys of the first and
fourth Galactic quadrants 
and fitted their data to an axisymmetric model of the space-averaged
density of the molecular gas as a function of Galactic radius, $R$, 
and height, $Z$, over the radial range 2~kpc$\, - R_\odot$. 
In Fig.~\ref{column_density}, we display their result for the radial dependence 
of the H$_2$ column density, $N_m(R)$, defined as the number of hydrogen nuclei 
tied up into H$_2$ molecules per unit area on the Galactic plane.
For completeness, the y-axis is labeled both in terms of $N_m(R)$
and in terms of the corresponding mass density per unit area, 
$\Sigma_m(R) = 1.42 \, m_{\rm P} \, N_m(R)$, where $m_{\rm P}$ 
is the proton rest mass.
Note that the strong central molecular peak, which falls outside 
the radial range explored by Bronfman \etalv does not appear on the figure.

Also shown in Fig.~\ref{column_density} is the azimuthally-averaged curve 
derived by Clemens \etal (1988) for the first Galactic quadrant.
This curve differs markedly from that obtained by Bronfman \etal (1988),
which, we recall, applies to the combined first and fourth quadrants:
it lies everywhere higher and has a more pronounced maximum at the location 
of the molecular ring.
Part of the difference reflects genuine large-scale departures from axisymmetry
and can be attributed to the existence of molecular spiral arms, which, 
due to their unwinding shape, cross the first and fourth quadrants 
at different Galactic radii.
However, the fact that Clemens \etals curve also differs from
Bronfman \etals fit to the first quadrant alone 
reveals a true discrepancy between both studies, which Bronfman \etal 
explained in terms of differences in instrumental calibrations, 
in statistical treatments, and in the adopted CO/H$_2$ ratio.

Beyond $R_\odot$, the H$_2$ column density averaged over azimuthal angle
drops off rapidly outward. To our knowledge,
there exists no quantitative estimate of its exact $R$-dependence,
except in restricted longitude intervals (e.g., Clemens \etalv 1988;
Grabelsky \etalv 1987), whose characteristics are not necessarily
representative of average properties along full Galactic circles.
The problem with partial azimuthal averages, already manifest 
in the inner Galaxy, probably becomes even worse beyond the solar circle, 
where arm-interarm density contrasts are more important.

Along the vertical, the molecular gas appears to be strongly confined 
to the Galactic plane, and its space-averaged distribution can be
approximated by a Gaussian. Upon averaging over azimuthal angle in the first
Galactic quadrant, Clemens \etal (1988) found that the full-width
at half-maximum (FWHM) of the molecular layer increases outward
as $R^{0.58}$ and has a value of $136 \pm 17$~pc at the solar circle. 
They pointed out that the observed thickening of the molecular layer 
with increasing radius is consistent with a decreasing stellar mass
surface density -- which entails a decreasing gravitational pull
toward the Galactic plane -- together with an approximately constant 
H$_2$ velocity dispersion along $Z$.
At $R_\odot$, they obtained for the space-averaged number density 
of hydrogen nuclei in molecular form
\begin{equation}
\langle n_m \rangle (Z) \, = \, \langle n_m \rangle (0) \
\exp \left[ - \left( {Z \over H_m} \right) ^2 \right] \ ,
\end{equation}
with $\langle n_m \rangle (0) = 0.58$~cm$^{-3}$ and $H_m = 81$~pc.
The axisymmetric model of Bronfman \etal (1988), fitted to the combined
first and fourth Galactic quadrant data, has a FWHM of $120 \pm 18$~pc,
roughly independent of $R$; at $R_\odot$, the molecular-hydrogen
space-averaged density is again given by Eq.~(1),
but with a midplane density $\langle n_m \rangle (0) = 0.53$~cm$^{-3}$
and a Gaussian scale height $H_m = 71$~pc.
The vertical profile of $\langle n_m \rangle$ or, equivalently,
that of the space-averaged molecular mass density, 
$\langle \rho_m \rangle = 1.42 \, m_{\rm P} \, \langle n_m \rangle$, 
is drawn in Fig.~\ref{space_density}, for both Bronfman \etals and
Clemens \etals studies.

High-resolution observations (see, for instance, Fig.~\ref{CO_map}) 
indicate that the molecular gas
is contained in discrete clouds organized hierarchically
from giant complexes (with a size of a few tens of parsecs,
a mass of up to $10^6 \ M_\odot$, 
and a mean hydrogen number density $\sim 100 - 1000$~cm$^{-3}$)
down to small dense cores (with a size of a few tenths of a parsec,
a mass $\sim 0.3 - 10^3 \ M_\odot$,
and a mean hydrogen number density $\sim 10^4 - 10^6$~cm$^{-3}$)
(Larson, 1981; Goldsmith, 1987).
The majority of molecular clouds are sufficiently massive to be bound
by self-gravity, and it can be verified that they approximately
satisfy the virial balance equation, $G \, M / R \sim \sigma^2$,
where $M$, $R$, and $\sigma$ are the cloud mass, radius, and internal
velocity dispersion, and $G$ is the gravitational constant
(Larson, 1981; Myers, 1987; but see also Maloney, 1990). 
They also roughly obey two empirical power-law relations,
$\sigma \propto R^{0.5}$ and $M \propto \sigma^4$, 
first obtained by Larson (1981) and confirmed by several subsequent
studies (e.g., Solomon \etalv 1987).
Because of the large scatter in the observational data points,
the normalization factors of these relations are ill-defined.
For reference, Solomon \etal (1987) found 
$\sigma_{\rm 1D} \simeq 1$~km~s$^{-1}$ and $M \simeq 2000~M_\odot$
for $R = 2$~pc.

From measurements of the peak specific intensity of CO emission lines,
it emerges that molecular clouds are, in general, extremely cold,
with typical temperatures in the range $10 - 20$~K (Goldsmith, 1987).
Thermal speeds at these low temperatures ($\sqrt{ 3 k T / m_{\rm H_2} }
\simeq 0.35 - 0.50$~km~s$^{-1}$) are small compared to the measured
internal velocity dispersions. This means that the total gas pressure
inside molecular clouds has but a small contribution from
its purely thermal component, the dominant contribution arising from
internal turbulent motions. 
Moreover, in accordance with the notion that molecular clouds are
gravitationally bound, the total gas pressure in their interior 
is much higher than in the intercloud medium.

H$_2$ molecules are believed to form by recombination of hydrogen atoms
on the surface of \is dust grains (Hollenbach and Salpeter, 1971).
The only regions where they can actually survive in vast numbers are 
the interiors of dark and translucent \is clouds (and possibly the deep
interior of diffuse clouds), which are simultaneously shielded from
radiative dissociation by external UV photons and cold enough to avoid
collisional dissociation (Shull and Beckwith, 1982).
The observed temperatures of molecular regions are easily explained 
as the result of thermal balance between heating by cosmic rays 
(and, at the cloud edges, collisions with photoelectrons from dust grains 
and with radiatively excited H$_2$ molecules)
and cooling by molecular line emission (primarily CO), the rate of which 
increases steeply with increasing temperature 
(de Jong \etalv 1980; Goldsmith, 1987; Hollenbach and Tielens, 1999). 
Collisions with dust grains also enter the thermal balance,
either as a coolant or as a heat source, depending on the dust temperature 
with respect to that of the gas (Burke and Hollenbach, 1983).

The main descriptive parameters of the molecular \is gas are listed
in Table~\ref{t1}.

\subsection{Neutral Atomic Gas}
\label{s4C}

Neutral atomic hydrogen, usually denoted by \hy (as opposed to \h2
for ionized hydrogen), is not directly observable at optical wavelengths. 
Under most \is conditions, particle collisions are so infrequent that 
nearly all hydrogen atoms have their electron in the ground energy level 
$n = 1$.
It turns out that all the electronic transitions between the ground
level and an excited state -- forming the Lyman series --
lie in the UV, with the Lyman $\alpha$ (L$\alpha$) transition
between the ground level and the first excited state $n = 2$
being at a wavelength of 1216~\AA.

Since its initial detection from a rocket-borne spectrograph 
(Morton, 1967), the \is L$\alpha$ line has been widely observed in
absorption against background stars and used to study the \hy
distribution in the local ISM. 
The method consists of aiming at a great number of nearby hot stars
distributed across the sky and analyzing their L$\alpha$ absorption line
to deduce the \hy column density between them and the Earth.

The early L$\alpha$ survey undertaken by Savage and Jenkins (1972) 
and extended by Jenkins and Savage (1974) showed that \hy is deficient
in the immediate vicinity of the Sun, especially in the third Galactic
quadrant (quadrant~III in Fig.~\ref{galaxy}).
We now understand the observed \hy deficiency as a consequence of 
the Sun's being located inside an \hy cavity, known as the Local Bubble
(Cox and Reynolds, 1987). This bubble is clearly asymmetric,
not only in the Galactic plane, where it extends significantly farther
in the longitude range $210^\circ - 250^\circ$,
but also along the vertical, where it reaches much higher altitudes 
in the northern hemisphere (e.g., Frisch and York, 1983).
Based on a compilation of L$\alpha$ absorption line measurements
by Fruscione \etal (1994), Breitschwerdt \etal (1996) estimated
that the \hy cavity has a radius in the plane $\sim 60 - 100$~pc
and a vertical extent from the plane $\sim 120 - 180$~pc.

A much more detailed and accurate outline of the \hy cavity was obtained
by Sfeir \etal (1999), who made use of the observational correlation
between the column density of neutral sodium (Na~{\sc i}) inferred from
its optical D-line doublet at 5890~\AA \ and the \hy column density 
inferred from the L$\alpha$ line (Hobbs, 1974).
They combined Na~{\sc i} column densities measured toward 456 stars
with the improved stellar distances provided by 
the {\it Hipparcos} satellite to draw contour maps of Na~{\sc i}
absorption near the Sun. According to these maps, the Local Bubble
has a radius in the plane varying between $\sim 60$~pc toward
Galactic longitude $l = 0$ and $\sim 250$~pc toward $l = 235^\circ$,
it is elongated along the vertical and possibly open-ended in the direction 
of the North Galactic Pole, and it is everywhere else surrounded by
a dense wall of neutral gas.

Following Savage and Jenkins' (1972) and Jenkins and Savage's (1974)
work, deeper, more reliable L$\alpha$ absorption surveys 
with the {\it Copernicus} satellite (Bohlin \etalv 1978) 
and with the {\it International Ultraviolet Explorer (IUE)} satellite
(Shull and Van Steenberg, 1985) made it possible to proceed with
methodical observations of the \hy gas outside the Local Bubble
and to gain a rough idea of its spatial distribution as a function of $Z$
in the solar neighborhood.
Unfortunately, the L$\alpha$ line as a diagnostic tool of \hy is plagued 
by the same \is extinction problem as UV and optical molecular lines,
which makes it unfit to map the \hy distribution at Galactic scales.
Here, too, one has to turn to radio astronomy.

The breakthrough event that opened the era of radio-astronomical 
observations of \is \hy was Ewen and Purcell's (1951) detection
of the \is 21-cm line emission predicted seven years earlier by
Hendrik van de Hulst.
The existence of the 21-cm line results from the ``hyperfine" structure
of the hydrogen atom. In brief, the interaction between the magnetic
moment of the electron and that of the proton leads to a splitting 
of the electronic ground level into two extremely close energy levels,
in which the electron spin is either parallel (upper level) or 
antiparallel (lower level) to the proton spin.
It is the ``spin-flip" transition between these two energy levels 
that corresponds to the now famous 21-cm line.
The major advantage of 21-cm photons resides in their ability 
to penetrate deep into the ISM, thereby offering a unique opportunity 
to probe the \is \hy gas out to the confines of the Milky Way.
On the other hand, the highly forbidden spin-flip transition 
is intrinsically so rare 
(Einstein $A$-coefficient $A_{21} = 2.85 \times 10^{-15}$~s$^{-1}$)
that very long paths are needed for the 21-cm line to be detectable.

In emission, the 21-cm line gives the total \hy column density
in the observed direction. Moreover, as explained in the Appendix, 
the contribution from each segment along the line of sight
can be extracted from the shape of the line profile combined with 
the Galactic rotation curve.
This is how 21-cm emission line measurements covering the whole sky
have been able to yield the \hy space-averaged density
as a function of position in the Galaxy.

\hy maps projected onto the Galactic plane exhibit long arc-like features 
organized into a spiral pattern (Oort \etalv 1958;
Mihalas and Binney, 1981, p.~528).
Overall, this spiral pattern appears rather complex and fragmented,
especially in the inner Galaxy ($R < R_\odot$) where the distance
ambiguity problem (see Appendix) largely contributes to confusing the picture.
Outside the solar circle, three major spiral arms clearly stand out 
in addition to the local minor Orion arm (Kulkarni \etalv 1982;
Henderson \etalv 1982). According to Kulkarni \etal (1982),
these arms have a roughly constant \hy surface density,
which is about four times greater than in the interarm regions.

Radially, the \hy gas extends out to at least 30~kpc from the Galactic center 
(Diplas and Savage, 1991). Its azimuthally-averaged column density
through the disk, $N_n(R)$, is characterized by a deep depression inside
3.5~kpc (Burton and Gordon, 1978), a relatively flat plateau through 
the solar circle (Lockman, 1984) and out to almost 14~kpc (Burton and
te Lintel Hekkert, 1986; Diplas and Savage, 1991), and an exponential 
fall-off beyond 14~kpc (Diplas and Savage, 1991). As a reminder, 
all the above lengths have been rescaled to $R_\odot = 8.5$~kpc.

For the outer Galaxy, Diplas and Savage's (1991) study, based on 
a single large-scale \hy survey, is limited to the longitude range 
$30^\circ - 250^\circ$, whereas both Henderson \etal (1982) and
Burton and te Lintel Hekkert (1986) managed to achieve almost full
longitudinal coverage by combining two complementary \hy surveys.
On the other hand, the \hy data analyzed by Henderson \etal (1982)
have a latitude cutoff at $\vert b \vert \leq 10^\circ$, 
which causes them to miss a significant fraction of the \hy gas, 
particularly at large Galactic radii.
Moreover, for technical reasons, the \hy density contours of
Henderson \etal (1982) and of Burton and te Lintel Hekkert (1986) 
become unreliable at a ten times higher density level than those of
Diplas and Savage (1991).
We, therefore, favor Diplas and Savage's (1991) results, which,
parenthetically, can be verified as consistent with Henderson \etals (1982) 
out to a radius $\sim 13$~kpc and with Burton and te Lintel Hekkert's (1986)
down to a density level $\sim 10^{-2}$~cm$^{-3}$.

In Fig.~\ref{column_density}, we plotted the composite function $N_n(R)$ 
constructed
by smoothly connecting a constant function at Dickey and Lockman's (1990)
best-estimate value of $N_n(R_\odot)$ (between 3.5~kpc and almost 14~kpc)
to a vanishing function (at small $R$) and
to an exponentially decreasing function averaged over the three
directions whose best-fit parameters were tabulated by Diplas and
Savage (1991) (outside 14~kpc). The averaged exponential function
should not be taken too seriously, both because the derived parameters
are quite different in the three fitted directions, thereby suggesting
that merely averaging over them does not lead to a trustworthy azimuthal 
average, 
and because the radial dependence of $N_n(R)$ outside the solar circle 
is very sensitive to the poorly known shape of the Galactic rotation curve.
Despite the important uncertainties at large $R$,
Fig.~\ref{column_density} underscores the stark contrast between 
the flat-topped profile of the \hy gas
and the peaked profile of the molecular gas.

The vertical structure of the \hy distribution is roughly uniform
for 3.5~kpc $< R < R_\odot$ (Lockman, 1984). In this radial interval,
the \hy gas lies in a flat layer with a FWHM of 230~pc (almost twice 
the FWHM of the molecular gas at $R_\odot$), 
and its space-averaged number density
can be approximated by the sum of two Gaussians and an exponential tail:
\begin{equation}
\langle n_n \rangle (Z) \, = \, (0.57 \ {\rm cm}^{-3}) \
\left\{ 
0.70 \ \exp \left[ - \left( {Z \over 127 \ {\rm pc}} \right) ^2 \right]
+ 0.19 \ \exp \left[ - \left( {Z \over 318 \ {\rm pc}} \right) ^2 \right]
+ 0.11 \ \exp \left( - {\vert Z \vert \over 403 \ {\rm pc}} \right)
\right\}
\end{equation}
(Dickey and Lockman, 1990; see Fig.~\ref{space_density}).
The thickness of the \hy layer drops to $\lapr 100$~pc inside 3.5~kpc
(Dickey and Lockman, 1990), and it grows more than linearly with $R$
outside $R_\odot$, reaching $\sim 3$~kpc at the outer Galactic boundary
(Diplas and Savage, 1991). This substantial flaring, expected from 
the steep decrease in the vertical gravitational field, 
is accompanied by a general warping of the \hy disk 
and by a regular scalloping of its outer edge (Kulkarni \etalv 1982;
Henderson \etalv 1982), whose physical origins are not well understood.
The warp is such that the midplane of the \hy layer lies above
the Galactic equatorial plane in the first and second quadrants,
with a maximum displacement $\sim 4$~kpc,
and below the Galactic plane in the third and fourth quadrants,
with a maximum displacement $\sim -1.5$~kpc 
(Dickey and Lockman, 1990; Diplas and Savage, 1991).
The scalloping has an azimuthal wavenumber $\simeq 10$
and an amplitude comparable to that of the warp (Kulkarni \etalv 1982).

21-cm absorption spectra generally look quite different from emission
spectra taken in a nearby direction: while the emission spectra contain 
both distinct narrow peaks and much broader features, only the narrow
peaks are present in the absorption spectra (see Fig.~\ref{HI_spectra}). 
The conventional interpretation of this difference is that the narrow
peaks seen in emission and in absorption are produced by discrete cold
($T \simeq 50 - 100$~K) \hy clouds, whereas the broader features seen 
in emission only are due to a widespread \hy gas that is too warm to give
rise to detectable 21-cm absorption.\footnote{
The pure-absorption coefficient is independent of temperature,
but the net absorption coefficient, corrected for stimulated emission, 
is inversely proportional to temperature.} 
The estimated temperature of the warm \hy component is 
$\simeq 6000 - 10000$~K (Dickey \etalv 1978; Kulkarni and Heiles, 1987).

Comparisons between 21-cm emission and absorption measurements indicate
that, in the vicinity of the Sun, the warm \hy has roughly the same
column density as the cold \hy (Falgarone and Lequeux, 1973; Liszt, 1983)
and about 1.5 times its scale height (Falgarone and Lequeux, 1973;
Crovisier, 1978).
The average fraction of cold \hy appears to remain approximately
constant from $R_\odot$ in to $\sim 5$~kpc and to drop by a factor
$\sim 2$ inside 5~kpc (Garwood and Dickey, 1989).
Outside the solar circle, \hy is probably mainly in the warm phase, 
as suggested by the fact that 21-cm emission profiles in the outer
portions of external face-on spiral galaxies are almost perfectly
Gaussian with a line width consistently in the range $6 - 9$~km~s$^{-1}$
(Dickey, 1996).

High-resolution maps of the 21-cm emission sky strikingly show that 
the cold \hy clouds are sheet-like or filamentary (Heiles, 1967; 
Verschuur, 1970; see also Fig.~\ref{HI_map}). 
Their true density can be estimated, for instance,
by measuring the relative populations of the three fine-structure levels
of the electronic ground state of \is neutral carbon (Jenkins \etalv 1983).
Typically, the hydrogen density in cold \hy clouds is found to be
$\simeq 20 - 50$~cm$^{-3}$, i.e., some two orders of magnitude larger than
in the warm intercloud \hy (Kulkarni and Heiles, 1987).
The fact that this density ratio is approximately the inverse of
the temperature ratio (see Table~\ref{t1}) supports the view that the cold 
and warm atomic phases of the ISM are in rough thermal pressure equilibrium.

The existence of two \hy phases with comparable thermal pressures 
but with radically different temperatures and densities was predicted 
theoretically by Field \etal (1969) (see also Goldsmith \etalv 1969), 
who demonstrated that atomic \is gas heated
by low-energy cosmic rays has two thermally stable phases:
a cold dense phase, in which the primary cooling mechanism 
is the radiative de-excitation of collisionally excited fine-structure
lines of metals, and a warm rarefied phase, resulting from the onset 
of L$\alpha$ cooling at about 8000~K.
Since Field \etals (1969) pioneering work, other heating mechanisms
have been put forward, such as photoelectric ejection off dust grains 
(Watson, 1972; Shull and Woods, 1985) and magnetohydrodynamic wave 
dissipation (Silk, 1975; Ferri\`ere \etalv 1988).
The presence of these additional heating mechanisms does not alter 
the general conclusion that cold and warm \is \hy may coexist in thermal
pressure balance (e.g., Wolfire \etalv 1995); this is a direct consequence
of the shape of the cooling curve, $\Lambda(T)$: fairly flat between
a steep rise due to the [C~{\sc ii}] 158~$\mu$m transition around 100~K 
and another steep rise due to L$\alpha$ around 8000~K.

Of course, the picture of a static ISM in strict equilibrium is
excessively idealized. The observed atomic clouds have 
random motions characterized by a one-dimensional velocity dispersion 
$\simeq 6.9$~km~s$^{-1}$ (Belfort and Crovisier, 1984).
A sizeable fraction of them
appear to be parts of expanding shells and supershells, with diameters 
ranging from a few tens of parsecs to $\sim 2$~kpc 
and with expansion velocities 
reaching a few tens of km~s$^{-1}$ (Heiles, 1979; Heiles, 1984).
Whereas the very most energetic supershells are thought to result from
the impact of infalling high-velocity clouds onto the Galactic disk,
all the more modest \hy shells are very likely created by stellar winds 
and supernova explosions, 
acting either individually or in groupings of up to a few thousands 
(Tenorio-Tagle and Bodenheimer, 1988).
Many of the clouds that do not seem to belong to any expanding shell
are probably fragments of old shells having lost their identity
(see Section~\ref{s6A}). Others may have directly condensed out of 
the warm neutral medium following a thermal instability induced
by converging gas motions (Hennebelle and P\'erault, 1999).

Let us note that the distinction between molecular and atomic 
clouds is not always clear-cut: some atomic clouds contain 
molecular cores (Dickey \etalv 1981; Crovisier \etalv 1984),
while many molecular clouds possess an atomic halo (Wannier \etalv 1983;
Falgarone and Puget, 1985). Moreover, much of the molecular material
and almost all the atomic material in the ISM share an important property
which ranks them amongst the ``photodissociation regions" (PDRs), namely, 
both are predominantly neutral and have their heating and chemistry
largely regulated by stellar UV photons (Hollenbach and Tielens, 1999).

\subsection{Warm Ionized Gas}
\label{s4D}

O and B stars, the most massive and hottest stars in the Milky Way, 
emit a strong UV radiation, which, below a wavelength of 912~\AA \
(corresponding to an energy of 13.6~eV),
is sufficiently energetic to ionize hydrogen atoms.
As a result, these stars are surrounded by a so-called ``\h2 region"
within which hydrogen is almost fully ionized.
Given that the ionizing UV photons are promptly absorbed by neutral hydrogen,
the transition between the \h2 region and the ambient ISM is rather abrupt.
Inside the \h2 region, ions and free electrons keep recombining before
being separated again by fresh UV photons from the central star.
Thus, the \h2 region grows until the rate of recombinations within it
becomes large enough to balance the rate of photoionizations.
In a uniform medium, this balance occurs when the radius of the \h2 region 
reaches the value of the Str\"omgren radius,
$$
r_S \, = \, (30 \ {\rm pc}) \
\left( {N_{48} \over n_{\rm H} \ n_e} \right) ^{1 \over 3} \ ,
$$
where $N_{48}$ is the number of ionizing photons emitted per unit time
by the central star, in $10^{48}$~s$^{-1}$ (e.g., $N_{48} \simeq 34$ 
for an O5V star and $N_{48} \simeq 1.7$ for a B0V star; Vacca \etalv 1996), 
and $n_{\rm H}$ and $n_e$ are the free-proton and free-electron
number densities in the \h2 region, in cm$^{-3}$ (Spitzer, 1978, p.~109).

The process of photoionization is accompanied by a net heating of the \is gas, 
as the ionizing photons transfer a fraction of their energy (the excess
with respect to the ionization potential) to the ejected electrons.
The equilibrium temperature, set by a balance between photoelectric 
heating and radiative cooling, has a typical value $\simeq 8000$~K,
depending on density and metallicity (Mallik, 1975; Osterbrock, 1989, p.~67).
This theoretical estimate turns out to be in good agreement with 
observational determinations based on measurements of the radio
continuum radiation (Osterbrock, 1989, p.~130) and on studies 
of emission line ratios (Osterbrock, 1989, p.~123) from \h2 regions.

The radio continuum radiation of an \h2 region arises from
the ``bremsstrahlung" or ``free-free" emission generated as free
electrons are accelerated in the Coulomb field of positive ions
(H$^{+}$, He$^{+}$, He$^{++}$).
Emission lines, found at optical, infrared, and radio wavelengths, 
are primarily due to radiative recombination of hydrogen and helium ions
with free electrons, and to radiative de-excitation of collisionally
excited ionized metals.
Of special importance are the optical hydrogen Balmer lines
produced by electronic transitions from an excited state $n > 2$ 
to the first excited state $n = 2$: because each recombination of a free
proton with a free electron into an excited hydrogen atom leads sooner
or later to the emission of one Balmer photon, 
and because the rate per unit volume of recombinations into an excited
hydrogen atom is $\propto n_{\rm H^{+}} \, n_e \propto n_e^2$,
the integrated intensity of the Balmer lines is directly proportional 
to the emission measure,
\begin{equation}
{\rm EM} \, = \, \int n_e^2 \ ds \ ,
\end{equation}
where $ds$ is the length element along the line of sight through 
the \h2 region.
For future reference, let us specify that the hydrogen Balmer
transition between the electronic energy levels $n = 3$ and $n = 2$
is usually referred to as the H$\alpha$ transition and has a wavelength
of 6563~\AA.

The presence of warm ionized \is gas outside well-defined \h2 regions
was first reported by Struve and Elvey (1938), who detected
H$\alpha$ and [O~{\sc ii}] 3727~\AA \ emission from extended zones
in Cygnus and Cepheus. Systematic studies of this gas, through its
optical emission lines, only started over thirty years later, with,
amongst others, an important H$\alpha$ photographic survey by Sivan (1974) 
and more sensitive H$\alpha$ spectroscopic scans by Roesler \etal (1978).
What emerged from these observational studies is that 
diffuse H$\alpha$-emitting gas exterior to \h2 regions exists in all
directions around us.
More recent, high-resolution H$\alpha$ maps of selected portions of the sky
display a complex structure made of patches, filaments, and loops of 
enhanced H$\alpha$ emission, superimposed on a fainter background
(Reynolds, 1987; Reynolds \etalv 1999a; see also Fig.~\ref{Halpha_map}).
The new {\it Wisconsin H$\alpha$ Mapper (WHAM)} survey
(Reynolds \etalv 1999a) will soon offer a complete and detailed view 
of the distribution and kinematics of the H$\alpha$-emitting gas 
over the entire sky north of $-30^\circ$.

The temperature of the diffuse emitting gas, inferred from the width of 
the H$\alpha$ and [S~{\sc ii}] 6716~\AA \ emission lines, is $\sim 8000$~K
(Reynolds, 1985a). This value has been confirmed by the recent {\it WHAM}
observations of H$\alpha$, [S~{\sc ii}] 6716~\AA \ and [N~{\sc ii}] 6583~\AA, 
which, in addition, suggest a temperature rise at high $\vert Z \vert$
(Haffner \etalv 1999). At 8000~K, the observed H$\alpha$ intensity
along the Galactic equator translates into an effective emission measure
in the range $9 - 23$~cm$^{-6}$~pc (Reynolds, 1983).
Combined with a mean free-path for absorption of H$\alpha$ photons 
in the Galactic disk $\simeq 2$~kpc (Reynolds, 1985b), this range
of emission measures implies a space-averaged electron density squared
$\langle n_e^2 \rangle \simeq 
4.5 \times 10^{-3} - 11.5 \times 10^{-3}$~cm$^{-6}$
in the diffuse ISM at low $\vert Z \vert$.
Across the Galactic disk, the emission measure is found to be 
$\sim 4.5$~cm$^{-6}$~pc (Reynolds, 1984), which, together with 
the above electron density squared yields an exponential scale height
$\sim 390 - 1000$~pc for the diffuse H$\alpha$ emission.

Owing to the obscuration effect of \is dust, the region that can be
probed with H$\alpha$ and other optical emission lines is limited
to a cylindrical volume of radius $\sim 2 - 3$~kpc around the Sun.
A totally different source of information on the warm ionized \is gas,
unaffected by obscuration, comes from the dispersion of pulsar 
(rapidly-spinning, magnetized neutron star, which emits 
regularly-spaced pulses of electromagnetic radiation) signals.
It is well known that electromagnetic waves travelling through an ionized
medium interact with the free electrons in such a manner that their group
velocity decreases with increasing wavelength.
The periodic pulses emitted by pulsars can each be decomposed
into a spectrum of electromagnetic waves spanning a whole range of radio
wavelengths, with the longer-wavelength waves propagating less rapidly
through \is space and, hence, arriving slightly later at the observer.
The resulting spread in arrival times, a measurable quantity,
is directly proportional to the column density of free electrons between
the pulsar and the observer, i.e., to the dispersion measure,
\begin{equation}
{\rm DM} \, = \, \int _0 ^L n_e \ ds \ ,
\end{equation}
with $L$ the distance to the pulsar.

Following the discovery of the first pulsar (Hewish \etalv 1968),
astronomers devised methods, mostly based on 21-cm absorption
measurements, to estimate pulsar distances.
This enabled them to model the large-scale distribution of \is free
electrons, by applying Eq.~(4) to pulsars with independent dispersion
measures and distance estimates.
The best-fit models contain a thin-disk component arising from localized
\h2 regions plus a thick-disk component associated with the diffuse warm
ionized medium (Manchester and Taylor, 1981; Harding and Harding, 1982;
Vivekanand and Narayan, 1982).
Near the Sun, the space-averaged density of free electrons
can be approximated by
\begin{equation}
\langle n_e \rangle (Z) \, = \, 
(0.015 \ {\rm cm}^{-3}) \ 
 \exp \left( - {\vert Z \vert \over 70 \ {\rm pc}} \right)
+ (0.025 \ {\rm cm}^{-3}) \ 
 \exp \left( - {\vert Z \vert \over 900 \ {\rm pc}} \right)
\end{equation}
(Reynolds, 1991), where the contribution from \h2 regions (first term)
is taken from Manchester and Taylor (1981), while the diffuse component
(second term) relies on the midplane density deduced from 
a limited sample of low-latitude pulsars by Weisberg \etal (1980) 
and on the column densities toward newly-discovered pulsars inside
high-$\vert Z \vert$ globular clusters.
As Reynolds (1991) himself admitted, the exponential scale height of 
the extended component in Eq.~(5) may have been underestimated by up to
a factor of 2, due to a probable deficiency inside the Local Bubble 
in which the Sun is located. 

Not only do pulsar signals experience dispersion upon propagating through
ionized regions, but they also get scattered by fluctuations in the 
free-electron density. A useful quantity in this context is the scattering
measure, SM, defined as the line-of-sight integral of the spectral
coefficient for a power-law spectrum of electron density fluctuations.
Scattering measures can be related to a number of observables, such as
the angular broadening of a small-diameter source and the temporal
broadening of pulsar pulses, and their observational determination
furnishes additional relevant information, particularly valuable
in the direction of the Galactic center and toward pulsars without
an independent distance estimate.

Cordes \etal (1991) analyzed two distinct data sets, comprising distances,
dispersion measures, and scattering measures of pulsars and other radio
sources, to construct an axisymmetric model of the free-electron
space-averaged density outside well-defined \h2 regions.
Their model consists of the superposition of a thin ($H_e = 150$~pc),
annular component centered on $R = 4$~kpc and a thick ($H_e = 1$~kpc),
radially extended component with Gaussian scale length $\gapr 20$~kpc.
The thin component, which is presumably linked to the molecular ring
discussed in Section~\ref{s4B}, gives a very small contribution 
at the solar circle, whereas the thick component corresponds to 
the second term in Eq.~(5) and approximately reduces to it at $R_\odot$.

Taylor and Cordes (1993) refined Cordes \etals (1991) model
by utilizing more extensive data sets and by allowing for departures
from axisymmetry. They explicitly incorporated a contribution from
spiral arms, based on the spiral pattern inferred from existing optical
and radio observations of \h2 regions, as well as a contribution from
the nearby Gum Nebula, which systematically enhances the dispersion
measure of pulsars located behind it. 
In their final model, the free-electron space-averaged density in 
interarm regions is somewhat less than in Cordes \etals (1991) 
axisymmetric model (for instance, at $R_\odot$,
$\langle n_e \rangle (Z \! = \! 0) = 0.019$~cm$^{-3}$ 
instead of $\simeq 0.025$~cm$^{-3}$); 
spiral arms add a contribution equal to 
$(0.08 \ {\rm cm}^{-3}) \ {\rm sech}^2 ( Z / 300 \ {\rm pc} )$
along their axis, so that, upon azimuthal average, both models are
in good agreement.

If we suppose that helium remains largely neutral in the warm ionized
medium, as suggested by the weak measured He~{\sc i} recombination line 
emission (Reynolds and Tufte, 1995; Tufte, 1997; Heiles \etalv 1996),
and if we disregard the fact that helium is fully ionized in the hot
medium (see Section~\ref{s4E}), we may identify the ionized-hydrogen
space-averaged density with the free-electron space-averaged density. 
Resorting to Cordes \etals (1991) axisymmetric model, we then obtain
the curves drawn in Figs.~\ref{column_density} and \ref{space_density} 
for the column density of ionized \is hydrogen as a function of $R$ 
and for its space-averaged density at $R_\odot$ as a function of $Z$, 
respectively.

Furthermore, if we assume a clear-cut separation between neutral and ionized 
media, with hydrogen completely neutral in the former and completely
ionized in the latter, and if we ignore the small fraction of free electrons 
arising in the hot medium, we can derive an estimate for the true density, 
$n_e$, and the volume filling factor, $\phi$, of the diffuse warm ionized medium
near the Sun. Indeed, from emission-measure data we know that 
$\langle n_e^2 \rangle \equiv \phi \ n_e^2 
\simeq 4.5 \times 10^{-3} - 11.5 \times 10^{-3}$~cm$^{-6}$.
On the other hand, dispersion-measure data yield $\langle n_e \rangle
\equiv \phi \ n_e \simeq 0.025$~cm$^{-3}$.
From this it follows that $n_e \simeq 0.18 - 0.46$~cm$^{-3}$
and $\phi \simeq 5 - 14~\%$ 
(repeating the reasoning of Kulkarni and Heiles, 1987).

The parameters of the diffuse warm ionized medium obtained in this subsection
are tabulated in Table~\ref{t1}. A comparison with the parameters of the cold
and warm neutral phases of the ISM suggests that the thermal pressure 
in the warm ionized medium ($\simeq 2.1 \, n \, k \, T$) is on average
roughly twice higher than in the neutral media 
($\simeq 1.1 \, n \, k \, T$).

Reynolds (1984) compared the \is hydrogen recombination rate inferred
from measurements of the Galactic H$\alpha$ emission to the ionizing power 
of known sources of ionizing radiation in the solar neighborhood,
and he concluded that only O stars are potentially able to do, 
by themselves, the desired job of maintaining the warm ionized medium 
in an almost fully ionized state (and, at the same time, 
at a temperature $\sim 8000$~K).
There exist, however, two inherent problems with O stars being 
the primary source of ionization.
First, O stars are preferentially born in dense molecular clouds 
close to the Galactic plane, which makes it difficult for a sufficient fraction 
of their ionizing photons to escape their immediate vicinity and pervade 
the general ISM up to the high altitudes where warm ionized gas is found 
(Reynolds, 1984).
Second, the observed emission-line spectrum of the diffuse ionized
background differs markedly from that characteristic of the compact 
\h2 regions surrounding O stars, with, in particular, an approximately
four times higher [S~{\sc ii}] 6716~\AA \ / H$\alpha$ intensity ratio
(Reynolds, 1985a). Moreover, the observed spatial variations of the 
[S~{\sc ii}] 6716~\AA \ / H$\alpha$ and [N~{\sc ii}] 6583~\AA \ / H$\alpha$
intensity ratios are difficult to explain by pure photoionization
(Reynolds \etalv 1999b).

The first problem can be overcome by taking into account 
the multi-component nature and the vertical structure of the ISM.
Adopting an ISM model consisting of a thin layer of small, opaque clouds
plus a more extended low-density extracloud medium, Miller and Cox (1993)
calculated the shape and size of the \h2 regions associated with the known 
O stars near the Sun; they showed that the most powerful of them were able 
to grow out of the cloud layer, up to high $\vert Z \vert$, and argued that
their dilute portions do, in fact, constitute the diffuse warm ionized
medium.
Another calculation by Dove and Shull (1994) suggests that the large,
elongated cavities blown by associations of O and B stars provide
natural channels for their ionizing photons to reach high-altitude regions
(see also Dove \etalv 2000).

To explain the high [S~{\sc ii}] / H$\alpha$ intensity ratio,
Sivan \etal (1986) had to appeal to a combination of photoionization
and weak-shock excitation, whereas Mathis (1986) and, later,
Domg\"orgen and Mathis (1994) found that a very dilute ionizing radiation 
field representative of a plausible mixture of O stars could reproduce 
the observations. 
The spatially variable [S~{\sc ii}] / H$\alpha$ and [N~{\sc ii}] / H$\alpha$ 
intensity ratios, for their part, appear to require supplemental 
ionization/heating sources (Reynolds \etalv 1999b), such as 
photoelectric ejection off dust grains (Reynolds and Cox, 1992), 
dissipation of \is plasma turbulence (Minter and Spangler, 1997), 
Coulomb encounters with Galactic \crs \ (Valinia and Marshall, 1998), 
or magnetic reconnection (Birk \etalv 1998).

\subsection{Hot Ionized Gas}
\label{s4E}

The notion that hot \is gas exists in the Milky Way dates back to
Spitzer's (1956) paper on a possible Galactic corona, made of hot
rarefied gas, which would provide the necessary pressure to confine 
the observed high-altitude \is clouds.
The presence of such a hot gas was born out almost two decades later 
by two independent types of observations:
(1) the {\it Copernicus} satellite detected, in the spectrum of several bright
stars, broad UV absorption lines of high-stage ions that form only
at elevated temperatures (Jenkins and Meloy, 1974; York, 1974),
and (2) the observed soft X-ray background radiation was found to be
most likely due to thermal emission from a hot \is plasma (Williamson
\etalv 1974).

Amongst the high-stage ions accessible to UV observations, 
O~{\sc vi} (five times ionized oxygen, with a doublet at (1032~\AA, 1038~\AA))
and N~{\sc v} (four times ionized nitrogen, with a doublet at (1239~\AA,
1243~\AA)) are the best tracers of hot collisionally ionized gas, 
insofar as their high ionization potential makes them difficult 
to produce by photoionization. Their degree of ionization together with
the measured line widths imply a temperature of a few $10^5$~K
(York, 1974; York, 1977). In addition, the integrated line intensities, 
which directly yield the column density of the considered ion between 
the Earth and the target stars, shed some light on its spatial 
distribution in the vicinity of the Sun.

The {\it Copernicus} O~{\sc vi} data were analyzed by Jenkins (1978a; 1978b), 
who, fitting the inferred O~{\sc vi} column densities with an exponential
along $Z$, arrived at a local scale height $\sim 300$~pc.
Unfortunately, this value is extremely uncertain, as most of the observed 
stars lie close to the Galactic plane.
Hurwitz and Bowyer (1996) studied more recent O~{\sc vi} data 
from a far-UV observation program of high-latitude stars and derived
an exponential scale height $\sim 600$~pc, which again should be taken 
with caution, given the wide fluctuations between sight lines.
Finally, Savage \etal (2000) measured O~{\sc vi} column densities
toward 11 active galactic nuclei (AGNs) with the {\it FUSE} satellite;
when coupling their measurements with a straight estimate of the O~{\sc vi}
midplane density from {\it Copernicus}, they obtained an O~{\sc vi}
exponential scale height of $2.7 \pm 0.4$~kpc, whereas by adopting
Shelton and Cox (1994) estimate of the O~{\sc vi} midplane density
-- which takes our position inside the Local Bubble into account --
they obtained a 35~\% higher scale height.

For N~{\sc v}, Sembach and Savage (1992) used multiple {\it IUE} absorption
spectra of a few bright halo stars to evaluate N~{\sc v} column densities
in several directions, and from their limited sample they deduced
an exponential scale height $\sim 1.6$~kpc.
With a larger set of stellar- and extragalactic-source absorption spectra
from both the {\it IUE} and the {\it Hubble Space Telescope (HST)},
Savage \etal (1997) came up with an N~{\sc v} exponential scale height 
of $3.9 \pm 1.4$~kpc. Note that these authors studied other high-stage ions 
likely to be related to the hot gas; by way of reference, they derived
exponential scale heights of $4.4 \pm 0.6$~kpc for C~{\sc iv}
and $5.1 \pm 0.7$~kpc for Si~{\sc iv}.

The soft X-ray background radiation around 0.25~keV appears to arise
predominantly from the Local Bubble (Cox and Reynolds, 1987). 
The temperature of the emitting gas, deduced from the relative
intensities of three adjacent energy bands, is $\simeq 10^6$~K
(McCammon and Sanders, 1990). Its average density can be inferred 
from the observed intensity of the soft X-ray flux, provided that 
the emission path lengths are known. Snowden \etal (1990) estimated
these path lengths by assuming that the X-ray--emitting region coincides
with the local \hy cavity and that the X-ray emissivity is uniform
throughout the cavity. The cavity's shape was then determined from 
the X-ray intensity distribution across the sky, and its scale was
adjusted such as to match at best 21-cm emission measurements of \hy
column densities. The resulting electron density is 
$\simeq 0.0037 - 0.0047$~cm$^{-3}$, which, if hydrogen and helium 
are fully ionized, 
corresponds to a hydrogen density $\simeq 0.0031 - 0.0039$~cm$^{-3}$.

After the 1990 launch of an X-ray telescope on the {\it R\"ontgen Satellite
(ROSAT)}, it became increasingly clear that a significant fraction of 
the 0.25-keV X-ray flux observed in a number of directions originates 
outside the Local Bubble, the best evidence
being the deep shadows cast in the soft X-ray background by \is clouds
in the Draco Nebula (Burrows and Mendenhall, 1991; Snowden \etalv 1991),
in the region of Ursa Major (Snowden \etalv 1994),
and in five other selected areas (Wang and Yu, 1995).
The discovery of these shadows led Snowden \etal (1998) to re-analyze
the 0.25-keV X-ray radiation, this time based on the high-resolution maps 
from the {\it ROSAT} all-sky survey, and model it with the superposition of 
an unabsorbed contribution from the Local Bubble, 
an absorbed contribution from the Galactic halo, 
and an absorbed isotropic contribution from extragalactic space.
Contrary to Snowden \etal (1990), they calibrated the bubble's scale
with the help of the estimated distance to a shadowing molecular cloud.
In this manner, they found a Local Bubble similar in shape, but somewhat
smaller than in the previous model, and they derived a hydrogen density
$\simeq 0.0065$~cm$^{-3}$.

If the values obtained for the temperature and the density of the hot gas 
in the Local Bubble are representative of the hot ionized phase of the ISM
near the Sun, the thermal pressure in the hot ionized phase 
($\simeq 2.3 \, n \, k \, T$) exceeds that in the neutral phases
($\simeq 1.1 \, n \, k \, T$) by a factor $\sim 3 - 15$ 
(see Table~\ref{t1}).

The 0.25-keV X-ray emission from the Galactic halo varies considerably
over the sky. The northern halo exhibits strong intensity enhancements
superimposed on a relatively uniform background, thereby suggesting that
the emitting gas has a patchy distribution. In contrast, the southern
halo is characterized by intensity gradients toward low latitudes, 
roughly consistent with a plane-parallel distribution (Snowden \etalv 1998).

It is very likely that 0.25-keV X-ray--emitting regions exist throughout
the Milky Way, but because their radiation is efficiently absorbed by
the intervening cool \is gas, the majority of them must escape detection.
On the other hand, a number of bright features have been observed in 
the intermediate energy band $0.5 - 1.0$~keV, which is less affected 
by photoelectric absorption. Most of these features were shown to be
associated either with individual supernova remnants (produced by 
isolated supernova explosions) or with ``superbubbles" (produced by 
the joint action of stellar winds and supernova explosions in a group 
of massive stars),
and their X-ray radiation was attributed to thermal emission from a hot
plasma at a temperature of a few $10^6$~K (Aschenbach, 1988;
McCammon and Sanders, 1990).

Present-day observations are too limited to enable astronomers to map
the large-scale distribution of hot \is gas in our Galaxy.
Nevertheless, some qualitative conclusions can be drawn from the
observational situation in external spiral galaxies. 
Their X-ray radiation is believed to arise from a combination of
unresolved discrete sources and diffuse thermal emission from hot \is gas 
(Fabbiano, 1989). Despite the difficulty to separate both contributions 
and to correct for line-of-sight obscuration by cool \is gas, Cui \etal
(1996) measured the soft X-ray intensity of several high-latitude
face-on spirals -- for which obscuration effects are minimal-- and derived
radial profiles for the emission measure of their hot component.
The derived profiles unambiguously show that the amount of hot gas
decreases radially outward. However, the large scatter among galaxies 
as well as the uncertainties involved in deducing column densities 
from emission measures and emission measures from measured intensities 
preclude any quantitative assessment.

It is now widely accepted that the hot \is gas is generated by 
supernova explosions and, to a lesser extent, by the generally powerful
winds from the progenitor stars (e.g., McCray and Snow, 1979; Spitzer, 1990).
Supernova explosions drive rapidly propagating shock waves in the ISM,
which sweep out cavities filled with hot rarefied gas and surrounded by
a cold dense shell of collapsed \is matter.
Cox and Smith (1974) pointed out that the hot gas inside the cavities 
had a sufficiently long radiative cooling time to be able to persist
for millions of years; they further argued that, for a Galactic
supernova frequency of one every 50~years, the hot cavities would
overlap and form a network of interconnecting tunnels. 
Elaborating on this idea, McKee and Ostriker (1977) developed
a self-consistent model of the local ISM, in which $\sim 70~\%$
of \is space turns out to be filled with hot gas. 
However, Slavin and Cox (1993) demonstrated that the \is 
magnetic pressure substantially reduces the hot gas filling factor 
(see also McKee, 1990). Another important issue is the tendency for
supernovae to be clustered and create superbubbles rather than
individual supernova remnants (McCray and Snow, 1979; Heiles, 1987).
Adopting a first crude model to describe the shape and temporal
evolution of superbubbles, Heiles (1990) estimated that they cover 
$\sim 17~\%$ of the Galactic disk area near the Sun.
With a more realistic model, including the latest observational
data on the ISM parameters and allowing for nearly equal numbers of
isolated and clustered supernovae, Ferri\`ere (1998b) obtained a local 
hot gas filling factor $\sim 20~\%$, with an overwhelming contribution
from superbubbles.

\subsection{Dust}
\label{s4F}

The most visible manifestation of \is dust,
already mentioned in Section~\ref{s2B}, is the obscuration and reddening 
of starlight as a result of absorption and scattering. 
The dust column density between a star and the Earth can be determined
observationally from the ``color excess" of the star, defined as
the difference between its measured color index and the intrinsic
color index given by its spectral type.
On the whole, the dust column density appears fairly closely correlated,
not with distance, but with hydrogen column density 
(Jenkins and Savage, 1974; Bohlin, 1975). This is a first indication that
\is dust tends to follow the inhomogeneous, patchy distribution
of the \is gas.

Extinction curves, which portray the wavelength dependence of \is
extinction toward individual stars, provide crucial clues to the nature
and size of the obscuring dust grains. These curves contain several
spectral features, which may be regarded as signatures of specific
radiative transitions. The fact that the relative strength of these
features varies from star to star strongly suggests that they are
produced by different types of grains (e.g., Meyer and Savage, 1981;
Witt \etalv 1984).
The most prominent feature, a bump at a UV wavelength $\simeq 2175$~\AA,
is traditionally attributed to graphite particles (Gilra, 1972; Mathis
\etalv 1977), although other carbon-containing compounds have also been
put forward (see Mathis, 1987, for a review).
The IR bands at 9.7~$\mu$m and 18~$\mu$m may be imputed to
amorphous silicates (Knacke and Thomson, 1973; Draine and Lee, 1984).
The carrier of a set of five mid-IR emission lines 
between 3.3 and 11.3~$\mu$m has been identified with polycyclic aromatic 
hydrocarbons (PAHs; Duley and Williams, 1981; L\'eger and Puget, 1984).
And other weaker features attest to the presence of additional species,
amongst which amorphous carbon and organic refractory material 
(Tielens and Allamandola, 1987).

The spectral shape of extinction curves may also be used to extract
information on the size distribution of \is dust grains, the upper end
of which may be further constrained by the cosmically available elemental
abundances. Mathis \etal (1977) were able to reproduce the standard
extinction curve of the diffuse ISM in the wavelength interval 
$0.1 - 1 \ \mu$m, with a mixture of spherical, uncoated graphite particles
and adjustable amounts of other substances (mainly silicates),
all distributed in size according to a power law of index $\simeq -3.5$,
$N(a) \, da \propto a^{-3.5} \, da$, over a radius range 
$a \simeq 0.005 - 1 \ \mu$m for the graphite particles
and $a \simeq 0.025 - 0.25 \ \mu$m for the other substances.
A similar size distribution, albeit with more structure between
0.02~$\mu$m and 0.2~$\mu$m and with a smooth fall-off between 0.2~$\mu$m
and 1~$\mu$m, was obtained by Kim \etal (1994), who relied on extinction
data extended to the wavelength interval $0.1 - 5 \ \mu$m and employed 
a more objective statistical treatment. Besides, Kim \etal (1994) confirmed
earlier claims that dense \is clouds have a flatter size distribution
than the diffuse ISM, with a greater proportion of large grains.

The energy contained in the stellar photons absorbed by the \is dust grains
heats them up to a temperature $\gapr 15$~K 
and is then mostly re-emitted in the IR.
Since the advent of IR astronomy in the early 1970's, it has been possible 
to observe the dust thermal emission 
and utilize it as an additional diagnostic tool to study the spatial
distribution, composition, and physical properties of \is dust.
Our knowledge in the field took two major steps forward with the experiments 
carried out by the {\it Infrared Astronomy Satellite (IRAS)} in 1983
and by the {\it Cosmic Background Explorer (COBE)} in $1989 - 1990$.
Both satellites returned all-sky maps of the diffuse IR emission,
the former with a finer spatial resolution and the latter with a broader
spectral coverage and better absolute calibration.

Analysis of {\it IRAS} maps showed that away from localized heating sources, 
there exists a good correlation
both between the 100-$\mu$m IR intensity outside molecular clouds and 
the \hy column density, and between the 100-$\mu$m IR intensity from
nearby molecular clouds and their H$_2$ column density
(Boulanger and P\'erault, 1988). 
IR emission measured by {\it COBE} at longer wavelengths is also well 
correlated with the \hy gas, and the excess emission observed in some 
regions can be explained by a fraction of the \is hydrogen being in molecular 
or ionized form (Boulanger \etalv 1996).
The best-to-date discussion of dust-gas correlations in the ISM 
was presented by Schlegel \etal (1998), who produced a composite all-sky map 
of 100-$\mu$m dust emission combining the {\it IRAS} spatial resolution
with the {\it COBE} quality calibration.

The \is dust emission spectrum covers a broad wavelength interval, 
from the near IR to the millimeter regime. With the help of preliminary 
{\it COBE} data, Wright \etal (1991) estimated for the first time 
the complete dust emission spectrum beyond 100~$\mu$m. 
Boulanger \etal (1996) later found that its long-wavelength portion 
is well fitted by a Planck function with temperature $\simeq 17.5$~K 
times an emissivity $\propto \lambda^{-2}$, 
although the emission peak near 150~$\mu$m 
and, more particularly, the excess emission below $\sim 100 \ \mu$m 
are incompatible with a single-temperature fit.

Independently of this observational finding, Draine and Anderson (1985)
had already shown that dust grains heated by ambient starlight
exhibit a whole range of temperatures, the width of which increases
with decreasing grain radius and becomes appreciable shortward of 0.01~$\mu$m.
This spread in temperatures results from the discrete nature of 
the heating process: because the typical energy of stellar photons 
is not small compared to the thermal energy of dust grains, 
each photon absorption causes a finite jump in the grain temperature, 
which subsequently cascades back down through successive emissions 
of IR photons.

Allowing for a realistic spread in grain temperatures makes it possible
to reproduce the observed dust emission spectrum.
For example, Dwek \etal (1997) obtained a very good fit up to a wavelength
$\simeq 500 \ \mu$m, with a dust model consisting of spherical graphite 
and silicate grains and planar PAH molecules, exposed to the local \is 
radiation field.
The graphite and silicate grains follow a slight variant of the power-law 
size distribution proposed by Mathis \etal (1977) (see above), 
while the PAH molecules are distributed in radius according to a power law 
of index $\simeq -3$ over a range $a \simeq 4 - 10$~\AA \
(assuming that the radius of a PAH, $a$, is related to 
its number of carbon atoms, $N_c$,
through $a = (0.913 \, {\rm \AA}) \ \sqrt{N_c}$; D\'esert \etalv 1990).
In Dwek \etals (1997) model, the far-IR emission beyond 140~$\mu$m
arises predominantly from the large graphite and silicate grains, 
whose temperature reaches $\simeq 17 - 20$~K and $\simeq 15 - 18$~K,
respectively, whereas the excess emission below 100~$\mu$m is due to
the small particles, which are stochastically heated to temperatures 
$\gg 20$~K (up to $\sim 500$~K for the PAHs).

The characteristics of \is dust grains, inferred from their IR emission
spectrum, vary among the different ISM phases. 
While the equilibrium dust temperature (temperature of the large grains) 
is relatively uniform at $\simeq 17.5$~K over most of the sky,
large-scale far-IR maps reveal the presence of cold ($\simeq 15$~K) dust,
statistically associated with molecular clouds (Lagache \etalv 1998).
Low dust temperatures (as low as 12~K) have also been observed 
in several dense condensations within nearby \is clouds 
(Ristorcelli \etalv 1998; Bernard \etalv 1999). 
In general, low dust temperatures appear to go on a par with
low abundances of small dust particles (Lagache \etalv 1998).
The dust emission associated with the warm ionized medium was first 
isolated by Lagache \etal (1999), whose preliminary analysis yielded 
a best-fit dust temperature $\simeq 29$~K for this medium.
Since then, the new results from the {\it WHAM} survey
have permitted a more trustworthy separation of the contribution
from the warm ionized medium to the observed dust emission,
which brought Lagache \etal (2000) to conclude that, in fact,
the dust temperature and abundances in the warm ionized medium 
do not differ significantly from those in \hy regions.

To make the link with the \is depletion factors discussed in Section~\ref{s4A},
let us note that in Dwek \etals (1997) dust emission model, 
silicate grains take up virtually all the cosmically available 
Mg, Si, and Fe, plus $\sim 15~\%$ of the available O,
graphite grains take up $\sim 62~\%$ of the available C, 
and PAHs take up another $\sim 18~\%$.
For comparison, Kim \etals (1994) extinction model, which does not
include PAHs, requires $\sim 95~\%$ of the available Si to be locked up
in silicate grains and $\sim 75~\%$ of the available C in graphite grains.
Hence direct studies of \is dust, based on its IR emission on the one hand
and on its extinction properties on the other hand, lead to similar
conclusions. These conclusions, however, are only partially supported
by \is depletion data, according to which the heavy refractory elements
are indeed highly depleted, but at least one-third of the carbon
remains in the gaseous phase (see Section~\ref{s4A}).

The origin of \is dust is not completely understood yet.
What seems firmly established is that a fraction of the dust grains 
form in the cool outer atmosphere of red-giant and supergiant stars
and in planetary nebulae,
where the temperature and pressure conditions are conducive to
condensation of carbonaceous species (graphite and amorphous carbon)
and silicates; the newly-formed particles are then expelled into \is space 
by the stellar radiation pressure (Woolf and Ney, 1969; Salpeter, 1976;
Draine, 1990).
On the other hand, two independent lines of evidence converge to suggest
that dust grains are also produced in the ISM itself:
On the theoretical side, the high-velocity shock waves driven by
supernova explosions were estimated to destroy dust grains 
at a much faster rate 
than the injection rate from stars, so that other sources are necessary 
to maintain the existing grain population (Dwek and Scalo, 1980;
Seab and Shull, 1983; Jones \etalv 1994).
On the observational side, the measured \is elemental depletions
tend to be higher in dense clouds (see Section~\ref{s4A}), which can
be explained by these cold entities being the sites of additional grain
formation\footnote{
Another possibility would be that low-density regions have statistically
undergone more grain destruction by fast shock waves (Shull, 1986).}
(Seab, 1987).

Although dust represents but a small fraction of the \is mass,
it plays a key role in the chemical and energetic balances of the ISM.
A first way in which dust participates in the ISM chemistry 
is by providing direct sinks and sources for the \is gas:
in the cold dense clouds, dust grains accrete particles from the gas phase,
whereas in the warmer intercloud medium, they shed their volatile mantle, 
which then returns to the gas phase; in addition, their refractory core 
is partially vaporized by each passing shock wave (Seab, 1987).
Another important chemical process to which dust grains largely contribute
is the formation and maintenance of molecular hydrogen: not only do they
serve as catalysts by allowing hydrogen atoms to recombine on their surface
(Hollenbach and Salpeter, 1971; Duley and Williams, 1993; Katz \etalv 1999), 
but they also help to shield the resulting H$_2$ molecules from 
photodissociation by the ambient UV radiation field (Shull and Beckwith, 1982).
Regarding the energetic balance of the ISM, dust gives a significant 
contribution to both heating and cooling, through ejection of energetic
photoelectrons (Watson, 1972; Bakes and Tielens, 1994)
and through collisions with gas particles (Burke and Hollenbach, 1983), 
respectively.

\section{Interstellar Magnetic Fields and Cosmic Rays}
\label{s5}

\subsection{Magnetic Fields}
\label{s5A}

The presence of \is \mfs \ in our Galaxy was first revealed by 
the observational discovery of linear polarization of starlight
(Hall, 1949; Hiltner, 1949a; Hiltner, 1949b).
Davis and Greenstein (1951) explained this polarization
in terms of selective extinction by elongated dust grains that are 
at least partially aligned by an \is \mf.
They showed that rapidly spinning paramagnetic grains tend to make 
their spin axis coincide with their short axis and orient the latter
along the \mf. Since dust grains preferentially block the component of light
with polarization vector ${\bf E}$ parallel to their long axis,
the light that passes through is linearly polarized in the direction of the \mf.
Applying their theory to the observed polarization of starlight, 
Davis and Greenstein (1951) already concluded that the \is \mf \
is locally parallel to the Galactic plane.

The first large-scale polarization data base 
was put together by Mathewson and Ford (1970), for a total of almost
7000 stars distributed over both celestial hemispheres and located
within a few kpc from the Sun.
Their compilation provides a comprehensive map of the \mf \ direction
on the plane of the sky, which not only substantiates Davis and
Greenstein's (1951) conclusions, but also indicates that the local \mf \
is nearly azimuthal, pointing toward a Galactic longitude $\simeq 80^\circ$,
i.e., having an inclination angle $\simeq 10^\circ$.
A recent, thorough re-analysis of Mathewson and Ford's (1970)
polarization data gives for the local \mf \ an inclination angle
$\simeq 7.2^\circ$ and a radius of curvature $\simeq 8.8$~kpc (Heiles, 1996);
this inclination angle of \mf \ lines is somewhat less than 
the standard pitch angle of the Galactic spiral pattern inferred from 
optical data ($\simeq 12^\circ$; Georgelin and Georgelin, 1976)
and from radio data ($\simeq 13^\circ$; Beuermann \etalv 1985).
Although considered the best reference for over a quarter-century,
Mathewson and Ford's (1970) catalog is now superseded by the more complete
compilation produced by Heiles (2000).

Stellar polarimetry acquaints us solely with the direction of the \is \mf.
To determine its strength, one has to resort to one out of three methods, 
based on the Zeeman splitting of the \hy 21-cm line or other radio lines, 
the Faraday rotation of linearly polarized radio signals,
and the radio synchrotron emission from relativistic electrons,
respectively .

The Zeeman splitting of a given atomic or molecular line occurs
in the presence of an external \mf, whose interaction with the magnetic
moment of the valence electrons causes a subdivision of certain
electronic energy levels. The amplitude of the Zeeman splitting,
$\Delta \nu$, is directly proportional to the \mf \ strength, $B$,
so that, in principle, it suffices to measure $\Delta \nu$ in order to
obtain $B$ in the region of interest. 
Unfortunately, in the case of the widely-used \hy 21-cm line,
$\Delta \nu$ is usually so small compared to the line width that 
it cannot be measured in practice. The way to circumvent this problem is
to observe instead the difference between the two circularly polarized
components of the 21-cm radiation, which directly yields the line-of-sight 
component of the \mf, $B_\parallel$.

The first successful implementation of this technique was performed by 
Verschuur (1969), who reported \mf \ strengths of a few $\mu$G
for several nearby \hy clouds.
Since Verschuur's early detections, a vast body of Zeeman-splitting
observations has built up, both for the \hy 21-cm line and for several
centimeter-wavelength lines of the OH molecule.
From a compilation of these observations by Troland and Heiles (1986),
it emerges that \is \mfs \ have typical strengths of a few $\mu$G
in regions with gas density $n \simeq 1 - 100$~cm$^{-3}$, and that 
they display only a slight tendency for $B$ to increase with increasing $n$;
this tendency is a little more pronounced in the higher density range 
$n \simeq 10^2 - 10^4$~cm$^{-3}$, where field strengths may reach up to
a few tens of $\mu$G (see also Myers \etalv 1995, Crutcher, 1999).
For comparison, the dipole \mf \ of the Earth has a strength of 0.31~G
at the equator, the solar \mf \ has a strength $\sim 1$~G in quiet regions
and $\sim 10^3$~G in sunspots, and pulsars have typical field strengths
$\sim 10^8 - 10^{13}$~G (e.g., Asseo and Sol, 1987).

Zeeman-splitting measurements are evidently biased toward \is regions
with high \hy column densities and narrow 21-cm line widths, i.e., toward
cold neutral clouds.
In contrast, Faraday-rotation measurements sample ionized regions.
As a reminder of basic plasma physics, a linearly polarized
electromagnetic wave propagating along the \mf \ of an ionized medium 
can be decomposed into two circularly polarized modes:
a right-hand mode, whose ${\bf E}$ vector rotates about the \mf \
in the same sense as the free electrons gyrate around it,
and a left-hand mode, whose ${\bf E}$ vector rotates in the opposite sense.
As a result of the interaction between the ${\bf E}$ vector and the free
electrons, the right-hand mode travels faster than the left-hand mode, 
and, consequently, the plane of linear polarization experiences a rotation 
-- known as Faraday rotation -- as the wave propagates.
The angle by which the polarization plane rotates is equal to 
the wavelength squared times the rotation measure,
\begin{equation}
{\rm RM} \, = \, {\cal C} \ \int _0 ^L n_e \ B_\parallel \ ds \ ,
\end{equation}
where the numerical constant is given by ${\cal C} = 0.81$~rad~m$^{-2}$
when the free-electron density, $n_e$, is expressed in cm$^{-3}$,
the line-of-sight component of the \mf, $B_\parallel$, is in $\mu$G,
and the path length, $L$, is in pc (Gardner and Whiteoak, 1966).
Observationally, the rotation measure of a given source can be 
determined by measuring the polarization position angle 
of the incoming radiation at two or more wavelengths.

The sources of linearly polarized waves used to carry out 
Faraday-rotation measurements in our Galaxy are either pulsars 
or extragalactic radio continuum sources.
Pulsars present a double advantage in this context: 
first, they lie within the Galaxy and have estimable distances,
and second, their rotation measure (Eq.~(6)) divided by their
dispersion measure (Eq.~(4)) directly yields the $n_e$-weighted average
value of $B_\parallel$ along their line of sight.

Rand and Kulkarni (1989) analyzed 116 pulsars closer than 3~kpc from the Sun
and concluded that the local \is \mf \ has a uniform (or regular) component
$\simeq 1.6 \ \mu$G and a random (or irregular) component $\sim 5 \ \mu$G.
By including pulsars deeper into the inner Galaxy and making a more
careful selection in the pulsar sample, Rand and Lyne (1994) arrived at
a local uniform field strength $\simeq 1.4 \ \mu$G. Moreover, they found that
the uniform field strength increases smoothly toward the Galactic center,
reaching at least 4.2~$\mu$G at $R = 4$~kpc (see their Fig.~6),
which implies an exponential scale length $\lapr 4.1$~kpc.
They also showed that a concentric-ring model, in which the uniform \mf \
is purely azimuthal, leads to field reversals at 0.4~kpc and at 3~kpc 
inward from the solar circle, with a maximum field strength between 
field reversals of 2.1~$\mu$G.

Han and Qiao (1994) obtained the same value $\simeq 1.4 \ \mu$G for the local
uniform field strength, but they argued that a bisymmetric \mf \ structure
with a pitch angle of $8.2^\circ$ and an amplitude of 1.8~$\mu$G
gives a better fit to the pulsar data than the concentric-ring model.
Han \etal (1999) found evidence for at least one field reversal in 
the outer Galaxy and possibly a third reversal inside the solar circle.
In line with most astrophysicists in the area, they claimed that 
the observed field reversals together with the measured pitch angle 
lend credence to the bisymmetic field picture.
Yet, it can be shown that field reversals are also consistent with
an axisymmetric magnetic configuration, and there exist reasons, 
both observational (Vall\'ee, 1996) and theoretical 
(Ferri\`ere and Schmitt, 2000), to favor the axisymmetric field picture.

Vertically, \is \mfs \ exist well beyond the pulsar zone, as indicated 
by the fact that the rotation measures of extragalactic radio sources
have the same sign and are systematically larger in absolute value 
than the rotation measures of pulsars in a nearby direction
(Simard-Normandin and Kronberg, 1980).
In principle, the scale height of the uniform field can be gathered from 
the observed rotation measures of high-latitude extragalactic sources
supplemented by a model of the free-electron density.
In practice, though, the inferred scale height turns out to be 
very sensitive to the parameters of the free-electron distribution,
which, regrettably, are poorly constrained.
For reference, from a sample of more than 600 extragalactic sources,
Inoue and Tabara (1981) obtained a magnetic scale height $\sim 1.4$~kpc.
This value is about one order of magnitude greater than the scale height
deduced from pulsar rotation measures (see Thomson and Nelson, 1980;
Han and Qiao, 1993). Despite the substantial uncertainties in both
estimates, the huge discrepancy could be indicative of the existence
of two magnetic layers with very different scale heights 
(Han and Qiao, 1994), a view supported by synchrotron-emission data 
(see Section~\ref{s5C}).

Regarding the \mf \ parity with respect to the Galactic midplane,
present-day observations do not convey a very clear picture.
The rotation-measure vertical distribution for extragalactic radio
sources and for pulsars with Galactic latitudes $\vert b \vert > 8^\circ$ 
appears to be approximately antisymmetric about the midplane 
in the inner (first and fourth) quadrants and roughly symmetric 
in the outer (second and third) quadrants 
(see Fig.~3 of Oren and Wolfe, 1995; Figs.~1 and 2 of Han \etalv 1997).
Although the observed antisymmetry in the inner Galaxy has often
been attributed to nearby anomalies like the North Polar Spur,
Han \etal (1997) argued for a genuine property of the uniform \mf \
away from the midplane.
Meanwhile, rotation measures of low-latitude pulsars point to
a symmetric distribution near the midplane, at all longitudes
(see Fig.~4 of Rand and Lyne, 1994).

To conclude this subsection, let us re-emphasize the two important
limitations of Faraday-rotation studies (see Heiles, 1995).
First, rotation measures essentially probe the warm ionized medium, 
which presumably contains most of the \is free electrons
but occupies only a modest fraction of the \is volume 
(see Section~\ref{s4D}), so that the inferred \mf \ strengths are not
necessarily representative of the ISM in general.
Second, rotation measures furnish the line-of-sight component of the 
Galactic \mf, not its total strength.
A more adequate method to determine \is \mf \ strengths draws on 
the Galactic radio synchrotron emission. Since the synchrotron
emissivity also depends on the density and spectrum of relativistic
electrons, we will return to this method in Section~\ref{s5C},
after discussing in detail the observed properties of \is \crs.

\subsection{Cosmic Rays}
\label{s5B}

The Earth is continually bombarded by highly energetic, electrically
charged particles from space.
Since their extraterrestrial origin was established by a balloon
experiment (Hess, 1919), they have been referred to as \crs,
even though it was later realized that they are in fact material
particles rather than photons (Bothe and Kohlh\"orster, 1929).
Measurements from instrumented balloons and satellites have shown that
\crs \ comprise protons, $\sim 10~\%$ of helium nuclei,
$\sim 1~\%$ of heavier nuclei, $\sim 2~\%$ of electrons,
and smaller amounts of positrons and antiprotons
(Bloemen, 1987; Blandford and Eichler, 1987).
They have typical velocities close to the speed of light
and span a whole range of kinetic energies, $E$.
While the majority of \crs \ with $E \lapr 0.1$~GeV/nucleon originate 
in the Sun, the more energetic ones emanate mainly from the ISM.

The observed \c_r energy spectrum is strongly modulated by the irregular, 
fluctuating \mf \ of the solar wind (Gleeson and Axford, 1968).
Because solar modulation effects are difficult to disentangle, 
the actual shape of the \is \c_r energy spectrum is uncertain below
$\sim 1$~GeV/nucleon.
At higher energies, its nuclear component can be approximated by
a piecewise continuous power law, $f(E) \propto E^{-\gamma}$,
with $\gamma \simeq 2.5$ for $E \simeq 2 - 10$~GeV/nucleon
and $\gamma \simeq 2.7$ for $E \simeq 10 - 10^5$~GeV/nucleon
(Simpson, 1983a; Webber, 1983a); the differential spectral index 
increases to $\gamma \simeq 3.1$ at $E \simeq 3 \times 10^6$~GeV/nucleon
and decreases again above $\sim 10^9$~GeV/nucleon (Hillas, 1984).
The electron spectrum runs parallel to the proton spectrum
between 2~GeV and 10~GeV and steepens to a differential index $\sim 3.3$
at a few tens of GeV (Webber, 1983b).

The energy density of \is \crs \ measured at Earth varies over
the course of the eleven-year magnetic cycle of the Sun,
from 0.98~eV~cm$^{-3}$ at sunspot minimum to 0.78~eV~cm$^{-3}$ 
at sunspot maximum. Webber (1987) extrapolated these values to 
the boundary of the heliosphere (cavity in the ISM carved out by 
the solar wind) by using the solar modulation model of 
Gleeson and Axford (1968) together with an estimate for their modulation
parameter based on the heliospheric \c_r intensity gradient
measured by the {\it Voyager} and {\it Pioneer} spacecraft 
out to $\sim 30$~AU from the Sun.\footnote{
The astronomical unit (AU) is the length unit equal to the distance 
between the Earth and the Sun. Its numerical value is given by 
1~AU = $1.50 \times 10^8$~km = $4.85 \times 10^{-6}$~pc.}
Proceeding in this manner, he obtained an \is \c_r energy density
$\simeq 1.5$~eV~cm$^{-3}$ just outside the heliosphere.
Eleven years later, with {\it Voyager} and {\it Pioneer} \c_r data available
out to $\gapr 60$~AU and with more sophisticated solar modulation models
(Potgieter, 1995) at his disposal, Webber (1998) updated the value of
the \is \c_r energy density to $\simeq 1.8$~eV~cm$^{-3}$.
At the time of this writing (October, 2000), {\it Voyager}~1 is at nearly 
80~AU from the Sun, probably quite close to the heliospheric termination shock, 
and Webber (private communication), who has periodically re-examined 
the situation, maintains the value of 1.8~eV~cm$^{-3}$.

If all \crs \ were ultrarelativistic, the \c_r pressure would simply
be one-third of their energy density, i.e., 
$\simeq 9.6 \times 10^{-13}$~dyne~cm$^{-2}$.
However, as re-emphasized by Boulares and Cox (1990),
the bulk of the \c_r energy density is due to mildly relativistic protons 
with kinetic energies of a few GeV, so the \c_r pressure must lie somewhere 
between one-third and two-thirds of their energy density. 
The correct pressure integral reads
$$
P_{\rm CR} \, = \, {1 \over 3} \ \int {E + 2 \, E_0 \over E + E_0} \
E \ n(E) \ dE \ ,
$$
where $n(E)$ is the \c_r differential number density and 
$E_0$ is the rest energy per nucleon (Ip and Axford, 1985).
Applying this expression to Webber's (1998) ``demodulated" \c_r spectrum
gives for the \is \c_r pressure just outside the heliosphere 
$(P_{\rm CR})_\odot \simeq 12.8 \times 10^{-13}$~dyne~cm$^{-2}$.

We know from absorption line studies toward the nearest (closer than 
a few parsecs) stars that the solar system is not directly surrounded by 
the hot tenuous gas of the Local Bubble, but that it is first immersed 
into a warm \is cloud, the Local Cloud, 
with a temperature $\simeq 6700 - 7600$~K, 
a \hy density $\simeq 0.18 - 0.28$~cm$^{-3}$,
and an electron density $\gapr 0.10$~cm$^{-3}$
(Frisch, 1995; Lallement \etalv 1996; Redfield and Linsky, 2000).
It is, therefore, reasonable to surmise that the aforementioned value 
of the \c_r pressure is representative of the warm phase of the ISM 
near the Sun.
The space-averaged value of $P_{\rm CR}$ near the Sun is probably
somewhat less, for magnetic and \c_r pressures are both expected to be
lower in the hot \is phase than in the rest of the ISM.  Indeed, supernova 
explosions, believed to constitute the main source of hot \is gas, 
expel \mf \ lines and \crs \ out of the hot cavities they create,
and it is only during their contraction phase that these hot cavities
are gradually replenished with field lines and \crs \ from the surrounding ISM.
Ferri\`ere (1998a) estimated that the space-averaged magnetic and \c_r
pressures near the Sun are only a few \% lower than their counterparts
in the warm Local Cloud. In view of the many uncertainties involved
in the above derivation, we will ignore this small difference
and ascribe the value
$(P_{\rm CR})_\odot \simeq 12.8 \times 10^{-13}$~dyne~cm$^{-2}$
to the space-averaged vicinity of the Sun.

The interaction of \crs \ with \is matter and photons gives rise
to $\gamma$-ray radiation through various mechanisms, including
(1) the production of $\pi^0$-mesons, which rapidly decay into two 
$\gamma$-photons,
(2) the Coulomb acceleration of \c_r electrons by the nuclei and
electrons of the \is gas, which leads to $\gamma$-ray bremsstrahlung
emission,
(3) the scattering of high-energy \c_r electrons on ambient soft photons,
which results in ``inverse-Compton" emission, again at $\gamma$-ray 
wavelengths (Bloemen, 1989).
Owing to technical difficulties, $\gamma$-ray astronomy had a slow start,
with several isolated balloon and satellite experiments, but since
the successful launch of the second {\it Small Astronomy Satellite (SAS-2)}
in 1972 (see Fichtel \etalv 1975) and the {\it Cosmic-Ray Satellite (COS-B)}
in 1975 (see Bloemen, 1989), rapid progress has been made in observing
the diffuse $\gamma$-ray background as a means to trace the Galactic 
distribution of \crs.
With a broader energy range ($\simeq 50 \ {\rm MeV} - 6 \ {\rm GeV}$) than 
{\it SAS-2} and with far better counting statistics due to its longer lifetime, 
the {\it COS-B} mission has been especially useful in this regard.

Correlation studies of {\it COS-B} $\gamma$-ray intensity maps 
with \hy and CO maps
show that the $\gamma$-ray emissivity per H-atom decreases away from
the Galactic center, this decrease being more pronounced at smaller
$\gamma$-ray energies (e.g., Bloemen \etalv 1986; Strong \etalv 1988).
The energy-dependence of the emissivity gradient was at first interpreted
as evidence that \c_r electrons  have a steeper radial gradient than 
\c_r nuclei (Bloemen \etalv 1986). However, follow-up work 
(Strong \etalv 1988) favored an alternative explanation, according 
to which molecular clouds, which are more numerous at small $R$, 
contain proportionally more \c_r electrons than the diffuse ISM.
In the second scenario, \c_r nuclei and electrons outside molecular clouds 
have the same radial dependence, with an exponential scale length
$\simeq 13$~kpc beyond 3.5~kpc (after rescaling to $R_\odot = 8.5$~kpc).

The available {\it COS-B} $\gamma$-ray maps convey little precise 
information on the vertical distribution of \is \crs,
first because the \is gas, which governs the $\pi^0$-decay and
bremsstrahlung components of the $\gamma$-ray emission,
tends to be confined close to the Galactic plane,
and second because the observed $\gamma$-ray radiation at Galactic
latitudes $\vert b \vert \gapr 10^\circ$ is probably contaminated 
by extragalactic sources.
The lower energy band $1 - 30$~MeV surveyed from the {\it Compton Gamma Ray
Observatory (CGRO)} is comparatively more sensitive to the inverse-Compton
component of the $\gamma$-ray emission. Attempts to isolate this component
suggest that it has significantly wider latitudinal extent than 
the \is gas (Strong \etalv 1996), which, if proven correct,  
implies that at least a fraction of the \c_r electrons are distributed 
in a thick disk.

It is possible to derive a more quantitative estimate of the \c_r 
vertical scale height on a totally different basis. 
In order to describe the method employed,
we first need to discuss the sources of \crs, their propagation
through the ISM, and their escape from the Galaxy.

The measured elemental composition of Galactic \crs \ points to two
different kinds of \c_r sources, both related to stars.
On the one hand, the similarity with the elemental composition of 
solar energetic particles, notably concerning an anti-correlation
between abundance and first ionization potential for elements heavier
than helium, suggests that Galactic \crs \ originate in unevolved
late-type stars, and are injected into the surrounding ISM
via flares out of their corona (Meyer, 1985).
On the other hand, the relative overabundance of iron (Simpson, 1983b)
argues for a formation of \crs \ in very evolved early-type stars,
with a release into the ISM upon the terminal supernova explosion.

Regardless of the exact injection sites, the injected \crs \
-- the so-called primaries -- are probably further accelerated upon
travelling in the ISM, through repeated scattering off moving
irregularities in the \is \mf.
This acceleration process may occur either stochastically
in the turbulent ISM (second-order Fermi acceleration; Fermi, 1949;
Fermi, 1954; Jokipii, 1978) or systematically at supernova shock waves,
where the converging upstream and downstream flows scatter \crs \
back and forth across the shock front (first-order Fermi acceleration;
Axford \etalv 1978; Bell, 1978; Blandford and Ostriker, 1978).

In spite of their relativistic velocities, \crs \ cannot freely travel
through \is space: they are trapped by the \is \mf, which constrains
their motion both perpendicular and parallel to its direction.
In the perpendicular direction, \crs \ are forced to gyrate about 
\mf \ lines along a circular orbit of radius
$$
r_g \, \simeq \, {E \over \vert q \vert \ B} \ \sin \theta
\, \simeq \, (10^{-6} \ {\rm pc}) \ 
{E({\rm GeV}) \over \vert Z \vert \ B(\mu{\rm G})} \ \sin \theta \ ,
$$
where $q$ is the \c_r electric charge, $Z$ is the atomic number
(taken as $-1$ for electrons), and $\theta = \arcsin (v_\perp / v)$
is the \c_r pitch angle, defined as the angle between its velocity 
and the \mf.
In the parallel direction, \crs \ excite resonant Alfv\'en waves,
which in turn scatter them and limit their streaming motion
to a slow diffusion at a velocity barely larger than the Alfv\'en speed
(Kulsrud and Pearce, 1969; Wentzel, 1969).
The literature abounds with theoretical models of \c_r
propagation and confinement in our Galaxy. We will not discuss these
models here, but we refer the interested reader to the comprehensive 
review paper by Cesarsky (1980).

What about the final fate of Galactic \crs?
A first possibility is that they end up losing all their energy 
in inelastic collisions with \is gas particles (Rasmussen and Peters, 1975).
A more attractive alternative is that they eventually escape from 
the Galaxy, either by streaming along \mf \ lines that connect with 
a weak extragalactic \mf \ (see Piddington, 1972), 
or by diffusing across field lines to the edge of the Galaxy, 
where they have a finite probability of ``leaking out" 
following magnetic reconnection
(e.g., in a magnetic bubble which detaches itself from 
the Galactic \mf; Jokipii and Parker, 1969),
or else by being convected away, together with the \is gas and 
field lines, in a Galactic wind (Jokipii, 1976).

Let us now return to the question of how Galactic \crs \ are distributed
along the vertical. When primary \c_r nuclei -- mainly C, N, and O --
collide with \is hydrogen, they can break up 
into lighter secondary nuclei -- such as Li, Be, and B.
The measured abundance of stable secondaries indicates that
the mean column density of \is matter traversed by \crs \
is energy-dependent and reaches a maximum value $\simeq 9$~g~cm$^{-2}$
at a \c_r energy $\simeq 1$~GeV/nucleon (Garcia-Munoz \etalv 1987).
Furthermore, the mean \c_r lifetime can be deduced from the measured abundance 
of unstable secondaries -- typically $^{10}$Be -- interpreted
in the framework of a \c_r propagation model.
If the volume occupied by \crs \ is modeled as a homogeneous ``leaky box",
the mean \c_r lifetime works out to be $\simeq 15$~Myr at 0.38~GeV/nucleon,
and the average ISM density in the \c_r box amounts to 
$\simeq 0.24$~cm$^{-3}$ (Simpson and Garcia-Munoz, 1988); 
since this is about one-fifth of the local ISM density, the \c_r box 
must be about five times thicker than the disk of \is gas.
In halo diffusion models, the mean \c_r lifetime is longer and
the effective thickness of the \c_r confinement region is larger 
than in the homogeneous leaky-box model.
For instance, Bloemen \etal (1993) derived an effective \c_r scale height 
$\lapr 3$~kpc at 1~GeV/nucleon. Likewise, Webber \etal (1992) found that
the thickness of the \c_r halo at 1~GeV/nucleon must lie between 2~kpc 
and $3 - 4$~kpc in the absence of Galactic wind convection, and that 
its maximum allowed value decreases with increasing wind velocity.

\subsection{Synchrotron Radiation}
\label{s5C}

The rapid spiraling motion of \c_r electrons about \mf \ lines 
generates nonthermal radiation, termed synchrotron radiation,
over a broad range of radio frequencies.
The synchrotron emissivity at frequency $\nu$ due to a power-law energy
spectrum of relativistic electrons, $f(E) = K_e \, E^{-\gamma}$,
is given by
\begin{equation}
{\cal E}_\nu \, = \, {\cal F}(\gamma) \ K_e \ B^{\gamma + 1 \over 2} \
 \nu^{- {\gamma - 1 \over 2}} \ ,
\end{equation}
where ${\cal F}(\gamma)$ is a known function of the electron spectral index 
and, as before, $B$ is the \mf \ strength (Ginzburg and Syrovatskii, 1965).
For the following, let us note that electrons with energy $E$ emit
the most power at frequency
\begin{equation}
\nu \, = \, (16 \ {\rm MHz}) \ B(\mu{\rm G}) \ E^2({\rm GeV}) 
\end{equation}
(Rockstroh and Webber, 1978).

Beuermann \etal (1985) modeled the Galactic synchrotron emissivity,
based on the all-sky radio continuum map at 408~MHz of Haslam \etal
(1981a; 1981b). Their model is restricted to the radial interval 
$3.5 - 17$~kpc (after rescaling to $R_\odot = 8.5$~kpc);
it possesses a spiral structure similar to that observed at optical 
wavelengths, and it contains a thin disk of equivalent half-thickness
$H_n(R) = (157~{\rm pc}) \ \exp \left[ (R - R_\odot) / R_\odot \right]$
plus a thick disk of equivalent half-thickness
$H_b(R) = (1530~{\rm pc}) \ \exp \left[ (R - R_\odot) / R_\odot \right]$.
The modeled synchrotron emissivity, averaged over Galactic azimuth,
can be written as
\begin{equation}
{\cal E}_\nu (R,Z) \, = \, (8.2 \ {\rm K} \ {\rm kpc}^{-1})
\left\{ 
0.46 \ \exp \left( - {R - R_\odot \over 2.8 \ {\rm kpc}} \right)
\left(  {\rm sech} \, {Z \over 255 \ {\rm pc}} \right)^{n(R)}
+ 0.54 \ \exp \left( - {R - R_\odot \over 3.3 \ {\rm kpc}} \right)
\left(  {\rm sech} \, {Z \over 255 \ {\rm pc}} \right)^{b(R)}
 \right\} \ ,
\end{equation}
where the exponents $n(R)$ and $b(R)$ take on the values imposed by
the above half-thicknesses; in particular, at the solar circle, 
$n(R_\odot) = 4.60$ and $b(R_\odot) = 0.187$.

Since the synchrotron emissivity (Eq.~(7)) depends on both the \mf \
strength and the spectrum of \c_r electrons, it is necessary to know
one of these two quantities in order to deduce the other one from Eq.~(9).
Unfortunately, as explained in Sections~\ref{s5A} and \ref{s5B},
neither quantity has been reliably determined so far.
One is thus led to follow another approach and appeal to a double
assumption often made in this context, namely, the assumption that 
the density of \c_r electrons is proportional to \c_r pressure
and that \c_r pressure is proportional to magnetic pressure. 
The second part of this double assumption may be justified by minimum-energy
type arguments (e.g., Beck \etalv 1996). 
The first part is certainly not strictly verified
throughout the electron spectrum, insofar as \c_r electrons suffer more
severe radiation losses (bremsstrahlung, inverse-Compton, synchrotron)
than \c_r nuclei. However, it is probably reasonable in the $2 - 10$~GeV 
energy range in which \c_r electrons generate most of the synchrotron
emission at 408~MHz (see below) and \c_r protons contribute most of 
the \c_r pressure (see Section~\ref{s5B}), since in that energy range 
the electron spectrum parallels the proton spectrum. 
Under these conditions, Eq.~(7) implies that magnetic pressure, 
$P_{\rm M} \equiv B^2 / 8 \pi$, and \c_r pressure, $P_{\rm CR}$, 
are both proportional to ${\cal E}_\nu^{4 \over \gamma + 5}$.

The last needed piece of information is the value of $P_{\rm M}$ and 
$P_{\rm CR}$ at a given point of the Galaxy, say, at the Sun.
The \c_r pressure at the Sun was estimated in Section~\ref{s5B} at
$(P_{\rm CR})_\odot \simeq 12.8 \times 10^{-13}$~dyne~cm$^{-2}$.
As far as magnetic pressure is concerned, there exist estimates 
from Zeeman-splitting and Faraday-rotation measurements, 
but these estimates refer only to the line-of-sight component 
of the \mf \ and, in addition, they are biased
toward cold neutral and warm ionized regions, respectively.
On the other hand, the local ISM constitutes a unique place in the Galaxy
for which it is possible to determine the \c_r electron spectrum
independently of other physical quantities, and, consequently,
to deduce the \mf \ strength from the value of the synchrotron emissivity, 
$({\cal E}_\nu)_\odot = 8.2~{\rm K}~{\rm kpc}^{-1}$ (see Eq.~(9)).

The \c_r electron spectrum in the Local Cloud harboring the solar system
can be directly measured at Earth down to an energy $E \sim 10$~GeV.
Below this energy, it becomes increasingly deformed by solar modulation
effects, but its shape is reflected in the shape of the synchrotron
emission spectrum, which is well established throughout the radio
frequency range 5~MHz $-$ 10~GHz.
Webber (1983b) was thus able to construct a composite \c_r electron
spectrum valid down to $E \sim 0.3$~GeV, by matching the unnormalized 
spectrum inferred from the synchrotron-emission spectral shape
to the normalized spectrum directly measured at Earth.

The \c_r electron spectrum derived by Webber (1983b) is not an exact
power law, so that the values of $\gamma$ and $K_e$ appearing
in Eq.~(7) vary weakly with electron energy.
The relevant energy here is the energy corresponding to $\nu = 408$~MHz
(the radio frequency at which the Galactic synchrotron emissivity 
has been modeled) and $B \simeq 5 \ \mu$G (the estimated \mf \ strength
near the Sun; see below), or, according to Eq.~(8), $E \simeq 2.3$~GeV.
At this energy, Webber's (1983b) composite spectrum yields $\gamma \simeq 2.5$
and $K_e \simeq (210$~m$^{-2}$~s$^{-1}$~sr$^{-1}$~GeV$^{-1}$) 
(2.3~GeV)$^{\gamma - 3}$.
Strictly speaking, the last value applies to the warm Local Cloud and
probably constitutes a slight overestimate for the space-averaged
vicinity of the Sun. We will, nonetheless, consider it as
the space-averaged value of $K_e$ near the Sun, 
for the same reasons as those invoked in Section~\ref{s5B},
when identifying the space-averaged value of $P_{\rm CR}$ near the Sun 
with its value in the Local Cloud.

We now introduce the above values of $\nu$, $\gamma$, and $K_e$,
together with $({\cal E}_\nu)_\odot = 8.2~{\rm K}~{\rm kpc}^{-1}$ 
(see Eq.~(9)), into Eq.~(7), whereupon we obtain for the local \mf \ strength
$B_\odot \simeq 5.0 \ \mu$G. The ensuing local magnetic pressure is
$(P_{\rm M})_\odot \simeq 10.0 \times 10^{-13}$~dyne~cm$^{-2}$.

In summary, we have derived the local values of the \is magnetic and
\c_r pressures, and we have argued that their large-scale spatial distribution
follows that of the synchrotron emissivity to the power
${4 \over \gamma + 5} \simeq 0.53$. 
Thus, if we denote by $F_{\cal E}(R,Z)$ the expression within curly
brackets in Eq.~(9), we may write
\begin{equation}
P_{\rm M}(R,Z) \, = \, (10.0 \times 10^{-13} \ {\rm dyne} \ {\rm cm}^{-2})
 \ \left[ F_{\cal E}(R,Z) \right]^{0.53}
\end{equation}
and
\begin{equation}
P_{\rm CR}(R,Z) \, = \, (12.8 \times 10^{-13} \ {\rm dyne} \ {\rm cm}^{-2})
 \ \left[ F_{\cal E}(R,Z) \right]^{0.53} \ \cdot
\end{equation}
For comparison, the thermal pressure inside the Local Cloud probably
lies in the range $2.8 - 6.8 \times 10^{-13}$~dyne~cm$^{-2}$ 
(Frisch, 1995; see Section~\ref{s5B}),
while that in the Local Bubble was recently estimated at
$\simeq 21 \times 10^{-13}$~dyne~cm$^{-2}$ 
(Snowden \etalv 1998; see Section~\ref{s4E}).
At the solar circle ($R = R_\odot$), Eqs.~(10) and (11) lead to 
the vertical profiles drawn in Fig.~\ref{pressure}.

According to Eq.~(10), the \is \mf \ strength, 
$B \propto \sqrt{P_{\rm M}} \,$,
has a radial scale length $\simeq 12$~kpc and a vertical scale height 
at the Sun $\simeq 4.5$~kpc, which are both significantly greater than
the values found for the uniform field component from Faraday-rotation
measurements (see Section~\ref{s5A}).
Vertically, one can argue that the \is \mf \ becomes less regular away
from the midplane (Boulares and Cox, 1990).
Radially, the discrepancy is more intriguing, as the outward increase 
in the degree of linear polarization of the synchrotron radiation from
external galaxies indicates that the regular field component falls off
with radius less rapidly than the total field strength (Heiles, 1995).
A possible explanation, offered by Heiles (1995), is that the ionized
regions sampled by Faraday rotation measures both have a weaker \mf \
than the neutral regions and occupy an outward-decreasing fraction of
the \is volume.

For Galactic \crs, Eq.~(11) predicts a radial scale length
$\simeq 6$~kpc, which is barely half that gathered from $\gamma$-ray
observations (see Section~\ref{s5B}).
Although part of the difference is likely due to uncertainties in the
modeled $\gamma$-ray and radio emissivities and in the fraction of these
emissivities truly attributable to \crs, an important source of disagreement
could reside in $\gamma$-ray observations being systematically biased
toward the dense ISM phases. In this respect, it is noteworthy that the \c_r 
scale length inferred from synchrotron measurements is easier to reconcile
with the short scale length of the presumed \c_r sources.
Finally, the \c_r equivalent scale height in Eq.~(11) is $\simeq 2$~kpc
at the solar circle, consistent with the maximum value $\simeq 3$~kpc
allowed by the diffusion-convection models of Bloemen \etal (1993) and
Webber \etal (1992).

\section{How everything fits together}
\label{s6}

Now that we have reviewed the different constituents of the ISM, let us
look into their interactions as well as their relations with stars.

\subsection{Role Played by Stars}
\label{s6A}

As already alluded to in the previous sections, stars affect the \is matter
essentially through their radiation field, their wind, and, in some
cases, their terminal supernova explosion. 
Globally, the massive, luminous O and B stars are by far the dominant players, 
even though they represent but a minor fraction of the stellar population 
(Abbott, 1982; Van Buren, 1985). 
Low-mass stars appear on the scene only for short periods of time,
during which they have important outflows or winds.

Stellar radiation photons, above all the energetic UV photons
from O and B stars, have a threefold direct impact on the \is matter.
(1) They dissociate H$_2$ molecules (provided $\lambda < 1120$~\AA) 
at the surface of molecular clouds (Federman \etalv 1979).
More generally, they dissociate molecules such as H$_2$, CO, OH, O$_2$,
H$_2$O $\ldots$ in photodissociation regions (Hollenbach and Tielens, 1999).
(2) They ionize the immediate vicinity of O and B stars,
thereby creating compact \h2 regions, and they ionize more remote 
diffuse areas, which together constitute the warm ionized medium
(see Section~\ref{s4D}).
In neutral regions, they ionize elements such as C, Mg, Si, and S, 
whose ionization potential lies below the 13.6-eV threshold of hydrogen 
(Kulkarni and Heiles, 1987).
(3) They heat up the \is regions that they ionize to a temperature
$\sim 8000$~K, by imparting an excess energy to the liberated
photoelectrons (see Section~\ref{s4D}). They also contribute to the heating 
of neutral regions, mainly through the ejection of photoelectrons from 
dust grains and through the radiative excitation of H$_2$ molecules followed by 
collisional de-excitation (see Sections~\ref{s4B}, \ref{s4C}, and \ref{s4F}). 
As a side effect of ionization and heating by stellar photons,
the traditional \h2 regions reach high thermal pressures,
which cause them to expand into the ambient ISM.

Stellar winds pertain to stars of all masses.
Low-mass stars are concerned only for limited periods in their lifetime.
Early on, just before joining the main sequence (prolonged phase of 
hydrogen burning in the stellar core), they experience energetic, 
more-or-less collimated outflows (Lada, 1985).
Toward the end of their life, after they have moved off the main
sequence, they successively pass through the red-giant, 
asymptotic-giant-branch (AGB), and planetary-nebula stages, 
during which they lose mass again at a very fast rate
(Salpeter, 1976; Knapp \etalv 1990).
High-mass stars suffer rapid mass loss throughout their lifetime
(Conti, 1978; Bieging, 1990). Their wind becomes increasingly powerful 
over the course of the main-sequence phase (e.g., Schaller \etalv 1992), 
and, if their initial mass exceeds $\simeq 32~M_\odot$, the wind reaches 
a climax during a brief post--main-sequence Wolf-Rayet phase
(Abbott and Conti, 1987).

Supernovae come into two types.
\sni \ arise from old, degenerate low-mass stars, which supposedly
are accreting from a companion and undergo a thermonuclear instability
upon accumulation of a critical mass.
\sn2 arise from young stars with initial mass $\gapr 8~M_\odot$,
whose core collapses gravitationally once it has exhausted all its fuel
(Woosley and Weaver, 1986).
Both types of supernovae release an amount of energy $\simeq 10^{51}$~ergs
(Chevalier, 1977).

To a large extent, stellar winds and supernova explosions act 
in qualitatively similar ways, although supernova explosions 
are more sudden and usually far more spectacular.
First of all, both constitute an important source of matter for the ISM.
Since this matter has been enriched in heavy elements by the thermonuclear 
reactions taking place inside the stars, the metallicity of the ISM 
is gradually enhanced.
The main contributors to the injection of mass into the ISM are the old
red-giant, AGB, and planetary-nebula stars (Salpeter, 1976; Knapp \etalv 1990).
As already mentioned in Section~\ref{s4F}, these cool stars are also 
outstanding in that their wind carries away newly-formed dust in addition 
to the regular gas.

Second, stellar winds and supernova explosions forge the structure 
of the ISM and are largely responsible for both its multi-phase nature
and its turbulent state. Here, the main contributors are the young,
massive O and B stars (Abbott, 1982; Van Buren, 1985).
To start with, the wind from a massive star blows a cavity of hot gas 
in the surrounding ISM and compresses the swept-up \is gas into 
a rapidly-expanding circumstellar shell 
(Castor \etalv 1975; Weaver \etalv 1977).
If it is initially more massive than $\simeq 8~M_\odot$, the star explodes
at the end of its lifetime (between $\simeq 3$~Myr for a 120-$M_\odot$
O3 star and $\simeq 38$~Myr for a 8-$M_\odot$ B3 star; Schaller \etalv 1992),
and the shock wave driven by the explosion pursues, in an amplified fashion, 
the action of the wind, sweeping up a lot more \is matter into the expanding 
shell, and greatly enlarging the hot cavity enclosed by the shell.
The compressed swept-up gas, at the elevated postshock pressure,
radiates efficiently, cools down, and collapses, so that the shell
soon becomes cold and dense (Woltjer, 1972; Chevalier, 1977;
Cioffi \etalv 1988).
Part of it may even turn molecular after typically $\sim 1$~Myr
(McCray and Kafatos, 1987).

If the shell collides with a comparatively massive \is cloud or,
at the latest, when the shock expansion velocity slows to roughly 
the external ``signal speed" (generalized sound speed, based on 
the total pressure, i.e., the gas + magnetic + cosmic-ray pressure, 
rather than the purely thermal pressure), 
the shell begins to break up and lose its identity.
The resulting shell fragments keep moving independently of each other
and start mixing with the \is clouds; at this point, the shell is said
to merge with the ambient ISM.
Meanwhile, the hot rarefied gas from the interior cavity comes into
contact with the ambient \is gas, mixes with it, and cools down
-- through thermal conduction followed by radiation --
to a temperature $\sim 10^4$~K.
Ultimately, what an isolated massive star leaves behind is a cavity
of hot rarefied gas, surrounded by an increasingly thick layer of warm
gas, plus several fragments of cold dense matter moving at velocities
$\sim 10$~km~s$^{-1}$. These fragments, be they atomic or molecular, 
appear to us as \is clouds.

The majority of O and B stars are not isolated, but grouped in clusters 
and associations (see catalog of Garmany and Stencel, 1992), 
so that their winds and supernova explosions act collectively to engender 
superbubbles (McCray and Snow, 1979; Heiles, 1987).
A superbubble behaves qualitatively like an individual supernova remnant,
with this difference that it has a continuous supply of energy.
For the first 3~Myr at least, this energy supply is exclusively due to 
stellar winds, whose cumulative power rises rapidly with time.
Supernovae start exploding after $\gapr 3$~Myr, and within $\sim 2$~Myr
they overpower the winds. From then on, the successive supernova
explosions continue to inject energy into the superbubble, at a slowly
decreasing rate, depending on the initial mass function of the progenitor 
stars, until $\sim 40$~Myr (McCray and Kafatos, 1987; Ferri\`ere, 1995). 
The time dependence of the energy deposition rate is in fact complicated
by the likely spread in star formation times, with deep implications for 
the superbubble's growth (Shull and Saken, 1995).
Altogether, stellar winds account for a fraction
comprised between $\sim 12~\%$ (Ferri\`ere, 1995) 
and $\sim 17~\%$ (Abbott, 1982) of the total energy input.

\sni \ are less frequent than their Type~II counterparts 
(see Section~\ref{s6B}).
All of them are uncorrelated in space, and they have basically the same 
repercussions on the ISM as isolated \sn2.

To fix ideas (see Ferri\`ere, 1998b), in the local ISM, the remnant of
a typical isolated supernova grows for $\sim 1.5$~Myr
and reaches a maximum radius $\sim 50$~pc. 
An ``average superbubble", produced by 30 clustered \sn2 
(see Section~\ref{s6B}), grows for $\sim 13$~Myr 
to a radius varying from $\sim 200$~pc in the Galactic plane 
to $\sim 300$~pc in the vertical direction.
This vertical elongation is a direct consequence of the ISM stratification:
because the \is density and pressure fall off away from the midplane, 
superbubbles encounter less resistance and, therefore, manage to expand
farther along the vertical than horizontally 
(see Tomisaka and Ikeuchi, 1986; Mac Low and McCray, 1988).

The hot gas created by supernova explosions and stellar winds 
in the Galactic disk rises into the halo under the effect of its buoyancy.
In the course of its upward motion, it cools down, almost adiabatically
at first, then through radiative losses, and eventually condenses 
into cold neutral clouds.
Once formed, these clouds fall back ballistically toward the Galactic plane.
The existence of such a convective cycle of \is matter between the disk
and the halo was first suggested by Shapiro and Field (1976), who dubbed it
``Galactic fountain". Detailed models of the Galactic fountain with
specific observationally-verifiable predictions were developed 
by Bregman (1980) and many other authors (see Bregman, 1996, for a review). 
Norman and Ikeuchi (1989) proposed a slight variant
in which the upward flow of hot gas concentrates in the ``chimneys"
formed by elongated superbubbles having broken out of the Galactic disk.
An attractive aspect of the Galactic fountain is that it furnishes 
a natural explanation for the observed existence of high-velocity clouds 
(\hy clouds whose measured velocity $\gapr 90$~km~s$^{-1}$ is too large 
to be solely due to the large-scale differential rotation) 
as well as for their measured velocity distribution, characterized
by a notable up-down asymmetry in favor of infalling clouds.
For the interested reader, the various possibilities for the origin 
of high-velocity clouds are reviewed by Wakker and van Woerden (1997).

Let us now inquire into the long-term evolution of the cold shell
fragments produced by supernova explosions. Some of them remain mostly
atomic and are observed as diffuse atomic clouds, moving randomly 
at velocities $\sim 10$~km~s$^{-1}$. 
Others, typically those arising from old superbubbles, become largely 
molecular, at least away from their surface.
These molecular fragments are responsive to self-gravity, which, 
past a critical threshold, drives them unstable to gravitational collapse.
The collapse of individual fragments or pieces thereof eventually leads
to the formation of new stars (e.g., Shu \etalv 1987), which, 
if sufficiently massive, may in turn initiate a new cycle of matter and 
energy through the ISM.

The idea of sequential OB star formation has been discussed by several
authors, including Mueller and Arnett (1976), Elmegreen and Lada (1977),
and McCray and Kafatos (1987). In particular, it has been suggested
that, once triggered in a molecular-cloud complex, OB star formation
could propagate and spread throughout the entire complex (Elmegreen and
Lada, 1977) and even possibly to more distant areas (Elmegreen, 1987).
However, as pointed out by McCray and Kafatos (1987),
self-induced star formation probably represents but a secondary process,
the primary triggering mechanism remaining, in all likelihood,
gas compression at the shock waves associated with the Galactic spiral arms 
(Lin and Shu, 1964).

On the other hand, the process of OB star formation in the Galaxy is,
to some extent, self-regulated. Norman and Silk (1980) had already
argued that outflows from pre--main-sequence low-mass stars
in a molecular cloud provide a continuous dynamic input, which maintains
the turbulent pressure at an adequate level to support the cloud against
gravitational collapse and, therefore, limit further star formation.
Franco and Shore (1984) applied a modified version of this argument 
to OB stars: through their radiation field, powerful wind, and supernova 
explosion, OB stars inject energy and momentum into their surrounding medium,
but because they do so much more vigorously than low-mass stars, 
they rapidly disrupt their parent molecular cloud and bring local star 
formation to a halt.
Other, milder regulatory mechanisms have been advocated, involving
either photoionization (McKee, 1989) or grain photoelectric heating
(Parravano, 1988).

Beside their obvious impact on the \is matter, stars are equally vital
for Galactic \crs \ and \mfs. As pointed out in Section~\ref{s5B},
stars are the likely birthplaces of most Galactic \crs,
and the shock waves sent by supernova explosions
constitute important sites of further \c_r acceleration.
Likewise, as we will touch upon in Section~\ref{s6C}, \is \mfs \ could
have their very first roots in stellar interiors; while this possibility
remains to be proven, there is now little doubt that, once a tiny \mf \
has been created, the turbulent motions generated by supernova explosions
amplify it at a fast rate.

\subsection{Supernova Parameters}
\label{s6B}

Given the particular importance of supernovae, we considered it necessary
to devote a full subsection to a description of their parameters.

The Galactic frequency of both types of supernovae can be estimated 
by monitoring their rate of occurrence in a large number of external 
galaxies similar to the Milky Way. From an eight-year observation program 
of 855 Shapley-Ames galaxies, Evans \etal (1989) derived average frequencies
of $0.2 \ h^2$~SNu for \sni \ and $1.7 \ h^2$~SNu for \sn2\footnote{
Type~Ib and Type~Ic supernovae, whose progenitors are now believed to be
young massive stars, were counted as \sn2.}
in Sbc-Sd galaxies, where $h$ is the Hubble constant in units of
100~km~s$^{-1}$~Mpc$^{-1}$ and 1~supernova unit (SNu) represents 
1~supernova per $10^{10} \, (L_B)_\odot$ per 100~yr.
More recently, Cappellaro \etal (1997) combined five independent
supernova searches (including Evans \etals (1989)), and from 
the resulting sample of 7773 galaxies, they derived average frequencies
of $0.41 \ h^2$~SNu for \sni \ and $1.69 \ h^2$~SNu for \sn2 in Sbc-Sc
galaxies.

The value of the Hubble constant, which gives the present expansion rate
of the Universe, was under heavy debate for over half a century, until
various kinds of observations made in the last few years finally converged 
to a narrow range $\simeq 60 - 70$~km~s$^{-1}$~Mpc$^{-1}$ (Branch, 1998;
Freedman \etalv 2001; Primack, 2000).
If we choose the median value of this range, corresponding to $h = 0.65$,
and assume that the Milky Way is an average Sbc galaxy with a blue
luminosity of $2.3 \times 10^{10} \, (L_B)_\odot$
(van den Bergh, 1988; van der Kruit, 1989), 
we find that Cappellaro \etals (1997) results translate into 
a Type~I supernova frequency
\begin{equation}
f_{\rm I} \, \simeq \, {1 \over 250~{\rm yr}}
\end{equation}
and a Type~II supernova frequency
\begin{equation}
f_{\rm II} \, \simeq \, {1 \over 60~{\rm yr}}
\end{equation}
in our Galaxy.

The corresponding total supernova frequency in our Galaxy is 
$\simeq 1 /$(48~yr), in reasonably good agreement with the evidence from
historical supernovae. Only five Galactic supernovae brighter than
zeroth magnitude were recorded in the last millenium, but it is clear
that many more supernovae occurred without being detected from Earth, 
mainly because they remained obscured by the \is dust.
Tammann \etal (1994) extrapolated from the five recorded events,
with the help of a detailed model of the Galaxy accounting for
obscuration by dust, and they concluded that a Galactic supernova frequency
$\sim 1 /$(26~yr) -- with a large uncertainty due to small-number
statistics -- could reproduce the historical observations.

The spatial distribution of supernovae is even more uncertain than their
frequency. In external spiral galaxies, both types tend to concentrate
to the spiral arms: like the bright massive stars, \sn2 are tightly
confined to the arms, whereas \sni \ have a more spread-out distribution,
similar to that of the general stellar population 
(McMillan and Ciardullo, 1996).
Radially, supernovae appear to be distributed with an exponential scale
length $\sim 2.4 - 3.8$~kpc for \sni \ and $\sim 3.0 - 5.5$~kpc for \sn2,
where the first and second values of each range refer to the regions 
$R < 5.8$~kpc and $R > 5.8$~kpc, respectively, and both values have been 
rescaled to a Hubble constant of 65~km~s$^{-1}$~Mpc$^{-1}$
(Bartunov \etalv 1992; see also van den Bergh, 1997).

It is also possible to infer the spatial distribution
of supernovae in our Galaxy from that of related objects.
For instance, we may reasonably suppose that \sni \ follow 
the distribution of old disk stars, with an exponential scale length 
$\simeq 4.5$~kpc along $R$ and an exponential scale height $\simeq 325$~pc
along $Z$ (Freeman, 1987). Adopting the Galactic frequency given by Eq.~(12), 
we may then write the Galactic Type~I supernova rate per unit area as
\begin{equation}
{\sigma}_{\rm I}(R) \, = \, (4.8 \ {\rm kpc}^{-2}~{\rm Myr}^{-1}) \
\exp \left( - {R - R_\odot \over 4.5 \ {\rm kpc}} \right)
\end{equation}
and their rate per unit volume at the solar circle as
\begin{equation}
{\cal R}_{\rm I}(Z) \, = \, (7.3 \ {\rm kpc}^{-3}~{\rm Myr}^{-1}) \
\exp \left( - {\vert Z \vert \over 325 \ {\rm pc}} \right) \ \cdot
\end{equation}

For \sn2, we may use either \h2 regions, which are produced by 
their luminous progenitor stars, or pulsars, which are the likely
leftovers of core-collapse explosions.\footnote{
In principle, one could also rely on the pulsar birthrate to evaluate 
the Type~II supernova frequency in our Galaxy.
Unfortunately, the pulsar birthrate is still poorly known.
It is, however, reassuring that its estimated value 
$\simeq 1 /$($30 - 120$~yr) (Lyne \etalv 1985)
or $\sim 1 /$(100~yr) (Narayan and Ostriker, 1990)
is compatible with Eq.~(13).}
McKee and Williams (1997) found that Galactic giant \h2 regions
are approximately distributed in a truncated exponential disk 
with a radial scale length $\simeq 3.3$~kpc over the radial range
$R \simeq 3 - 11$~kpc.
Galactic pulsars, for their part, were shown to be radially distributed
according to a rising Gaussian with a scale length $\simeq 2.1$~kpc 
for $R < 3.7$~kpc and a standard Gaussian with a scale length 
$\simeq 6.8$~kpc for $R > 3.7$~kpc (Narayan, 1987; Johnston, 1994).
Their vertical distribution at birth can be approximated by 
the superposition of a thin Gaussian disk with a scale height
$\simeq 212$~pc and a thick Gaussian disk with three times the same
scale height, containing, respectively, 55~\% and 45~\% of the pulsar
population (Narayan and Ostriker, 1990).
In view of Eq.(13), the pulsar model leads to a Galactic Type~II
supernova rate per unit area
\begin{equation}
{\sigma}_{\rm II}(R) \, = \, (27 \ {\rm kpc}^{-2}~{\rm Myr}^{-1}) \
\left\{
\begin{array}{ll}
3.55 \ \exp \left[ - \left( { \displaystyle R - 3.7 \ {\rm kpc} 
                              \over
                              \displaystyle 2.1 \ {\rm kpc} } 
                     \right)^2 \right] \ , \qquad
& R < 3.7 \ {\rm kpc}
\nonumber  \\
\noalign{\medskip}
\exp \left[ - { \displaystyle R^2 - R_\odot^2 
                \over 
                \displaystyle (6.8 \ {\rm kpc})^2 } \right] \ , \qquad
& R > 3.7 \ {\rm kpc}
\end{array}
\right.
\end{equation}
and a Type~II supernova rate per unit volume at $R_\odot$
\begin{equation}
{\cal R}_{\rm II}(Z) \, = \, (50 \ {\rm kpc}^{-3}~{\rm Myr}^{-1}) \
\left\{ 
0.79 \ \exp \left[ - \left( {Z \over 212 \ {\rm pc}} \right)^2 \right] 
+ 0.21 \ \exp \left[ - \left( {Z \over 636 \ {\rm pc}} \right)^2 \right]
\right\} \ \cdot
\end{equation}
Note that the mean height predicted by Eq.~(17) is significantly larger
than the mean height of OB stars ($\simeq 90$~pc; Miller and Scalo, 1979),
consistent with the fact that these stars are usually born close to 
the midplane and progressively increase their average distance from it
as they grow older.
The Galactic supernova rates per unit area and per unit volume are plotted 
in Figs.~\ref{rate_area} and \ref{rate_volume}, respectively, 
for both types of supernovae.

Before closing this subsection, let us estimate the fraction of \sn2
that are clustered and the way they are distributed amongst different 
clusters. In the catalog of 195 Galactic O stars compiled by Gies (1987),
71~\% lie in groups and 29~\% lie in the field.
For the O stars in groups, we may use the radial peculiar velocities
tabulated by Gies together with the assumption of isotropy in peculiar-velocity
space to reconstruct the distribution in total peculiar velocity.
This distribution clearly possesses an excess of high-velocity stars,
amounting to $\simeq 15~\%$ of the group stars, i.e., $\simeq 11~\%$ of all 
O stars. According to Gies' interpretation, these high-velocity stars were
recently ejected from their native cluster and will end up in the field. 
From this, we conclude that $\sim 60~\%$ of the O stars were born and 
will remain in groupings, while $\sim 40~\%$ of them will die in the field.

Can these figures be extended to all Type~II supernova progenitors?
Humphreys and McElroy (1984) compiled a list of all known-to-date
Galactic luminous stars and found that 47~\% of them are grouped.
Since their list contains on average older stars than Gies' (1987),
it is not surprising that a larger fraction of their stars appear 
in the field. Indeed, stars born in a group may after some time be
observed in the field, either because they have been ejected from 
the group or because the group has dispersed. If we accept that 
$\simeq 11~\%$ of the O stars are observed as group members but will
end up in the field as a result of ejection, and if we assume that OB
associations disperse into the field after $\sim 15$~Myr
(Blaauw, 1964; Mihalas and Binney, 1981, p.~164),
we find that Humphreys and McElroy's (1984) compilation is consistent 
with 60~\% of all \sn2 being clustered.

Clustered \sn2 are very unevenly divided between superbubbles.
In other words, the number of clustered supernovae contributing to
one superbubble, $N$, is extremely variable. 
The distribution of $N$ can be deduced from the observed
luminosity distribution of \h2 regions, either in external galaxies
(Kennicutt \etalv 1989; Heiles, 1990) or in our own Galaxy 
(McKee and Williams, 1997).
Both kinds of observations suggest a power-law distribution in $N^{-2}$.
Moreover, relying on local observations of OB stars and stellar clusters,
Ferri\`ere (1995) estimated that $N$ averages to $\simeq 30$ 
(in agreement with an earlier estimation by Heiles, 1987) 
and varies roughly between 4 and $\sim 7000$.

\subsection{Role Played by Cosmic Rays and Magnetic Fields}
\label{s6C}

Interstellar \crs, these extremely energetic and electrically charged
particles pervading \is space, impinge on the \is matter in three important
ways.
(1) They contribute to its ionization, through direct collisions with
\is gas particles (Spitzer and Tomasko, 1968).
(2) They constitute a triple source of heating, arising from the excess 
energy carried away by the electrons released in \c_r ionization
(Field \etalv 1969), from Coulomb encounters with charged particles 
of the ordinary gas (Field \etalv 1969), and from the damping of Alfv\'en 
waves excited by cosmic rays' streaming along \mf \ lines (Wentzel, 1971).
(3) They are dynamically coupled to the \is matter via the intermediary
of the \is \mf, more specifically, by their gyromotion in the perpendicular 
direction, and by their scattering off self-excited Alfv\'en waves 
in the parallel direction (see Section~\ref{s5B}).
In consequence, they exert their full pressure on the \is matter,
thereby affecting its dynamics.

The \is \mf \ acts on the \is matter through the Lorentz force.
Of course, the field acts directly on the charged particles only, 
but its effect is then transmitted to the neutrals by ion-neutral
collisions (Spitzer, 1958).
Apart from the densest parts of molecular clouds, whose ionization
degree is exceedingly low ($x \sim 10^{-8} - 10^{-6}$; Shu \etalv 1987),
virtually all \is regions are sufficiently ionized
($x \sim 4 \times 10^{-4} - 10^{-3}$ in cold atomic clouds
and $x \sim 0.007 - 0.05$ in the warm atomic medium;
Kulkarni and Heiles, 1987) for their neutral component to remain tightly
coupled to the charged component and, hence, to the \mf.

At large scales, the \is \mf \ helps to support the ordinary matter
against its own weight in the Galactic gravitational potential,
and it confines \crs \ to the Galactic disk.
In this manner, both \mfs \ and \crs \ partake in the overall
hydrostatic balance of the ISM and influence its stability.
Boulares and Cox (1990) were the first authors to fully appreciate
the importance of \mfs \ and \crs \ in the hydrostatic balance.
By the same token, they managed to solve the long-standing problem
of apparent mismatch between the total \is pressure at a given point
and the integrated weight of overlying \is material: by adopting
higher magnetic and \c_r pressures than previously estimated, they were
able to bring the total pressure at low $\vert Z \vert$ into agreement
with the integrated weight, and by including the magnetic tension force
at high $\vert Z \vert$, they could explain why the weight integral
falls off faster than the total pressure.

As magnetic and \c_r pressures inflate the gaseous disk, they tend 
to make it unstable to a generalized Rayleigh-Taylor instability, 
now known in the astrophysical community as the Parker instability 
(Parker, 1966; see also Zweibel, 1987).
When this instability develops, \mf \ lines ripple, and the \is matter
slides down along them toward the magnetic troughs, where it accumulates.
This whole process, it has been suggested, could give birth to new
molecular-cloud complexes and ultimately trigger star formation
(Mouschovias \etalv 1974; Elmegreen, 1982).

At smaller scales, the \is \mf \ affects all kinds of turbulent motions
in the ISM. Of special importance is its impact on supernova remnants
and superbubbles (see Tomisaka, 1990; Ferri\`ere \etalv 1991; 
Slavin and Cox, 1992).
First, the background magnetic pressure acting on their surrounding
shells directly opposes their expansion.
Second, the magnetic tension in the swept field lines gives rise to 
an inward restoring force, while the associated magnetic pressure 
prevents the shells from fully collapsing and, therefore, keeps them 
relatively thick.
Third, the enhanced external ``signal speed" causes the shells to merge
earlier than they would in an unmagnetized medium.
All three effects conspire to lower the filling factor of hot cavities
(Ferri\`ere \etalv 1991; Slavin and Cox, 1993).

The \is \mf \ also constrains the random motions of \is clouds 
which are not, or no longer, parts of coherent shells.
The basic physical idea is the following:
Interstellar clouds are magnetically connected to their environment, 
namely, to the intercloud medium and possibly to neighboring clouds,
through the \mf \ lines that thread them. When a given cloud moves
relative to its environment, these field lines get deformed, 
and the resulting magnetic tension force modifies the cloud's motion,
transferring part of its momentum to its environment (Elmegreen, 1981).
Likewise, angular momentum can be transferred from a cloud to
its environment by magnetic torques (Mouschovias, 1979;
Mouschovias and Paleologou, 1979).
The latter mechanism is particularly relevant to the star formation process, 
as it allows the contracting protostellar cores to get rid of angular 
momentum (e.g., Mouschovias and Morton, 1985). 

Finally, \mfs \ play a crucial role in the support of molecular clouds
against their self-gravity and in the eventual gravitational collapse 
of protostellar cores.
The magnetic support of molecular clouds is essentially provided by
magnetic pressure gradients in the directions perpendicular to 
the average field and, presumably, by nonlinear Alfv\'en waves
in the parallel direction (see review paper by Shu \etalv 1987).\footnote{
Beside the direct, stabilizing effect of \mfs, there exists an indirect,
possibly destabilizing effect, due to the distortion of the equilibrium
configuration (Zweibel, 1990).}
In the case of protostellar cores, magnetic support is insufficient 
to prevent their ultimate gravitational collapse.
This is generally because their ionization degree is so low that
neutrals are not perfectly tied to \mf \ lines, which enables them
to drift inwards under the pull of their self-gravity and eventually
form stars (Nakano, 1979; Mestel, 1985).

The last issue we would like to discuss pertains to the very origin 
of \is \mfs. The most likely scenario to date involves a hydromagnetic
dynamo, based on the concept that the motions of a conducting fluid embedded
in a \mf \ generate electric currents, which, under favorable conditions,
can amplify the original \mf.
In the Galaxy, the dynamo process appeals to a combination of
large-scale differential rotation and small-scale turbulent motions
(Steenbeck \etalv 1966; Parker, 1971; Vainshtein and Ruzmaikin, 1971).
Remember that the \is gas is tied to \mf \ lines, or, stated more
appropriately here, \mf \ lines are ``frozen-in" into the \is gas.
Accordingly, the large-scale differential rotation stretches field lines
in the azimuthal direction about the Galactic center, thereby
amplifying the azimuthal component of the large-scale \mf.
Meanwhile, its radial component is amplified by small-scale turbulent
motions, through a mechanism called the ``alpha-effect", the principle
of which can be described as follows: 
turbulent motions, taking place in a rotating medium,
are acted upon by the Coriolis force, which imparts to them
a preferred sense of rotation; in consequence, the small-scale magnetic 
loops that they produce are preferentially twisted in a given sense, 
and the net result is the creation of mean \mf \ in the direction
perpendicular to the prevailing field.
In addition to being responsible for the alpha-effect, turbulent motions
also contribute to the vertical escape of \mf \ lines and to their
spatial diffusion.

It is important to realize that the operation of the Galactic dynamo
requires a seed magnetic field to initiate the amplification process. 
Several possibilities have been advanced regarding the nature
and origin of this seed field (see Rees, 1987, for a review).
Briefly, the seed field could be an extragalactic \mf \ already present 
in the Universe prior to galaxy formation, or it could arise 
in the protogalaxy as a result of charge separation due to electrons
interacting with the microwave background photons, or else it could
originate in the first generation of stars and be expelled into the ISM
by their winds and/or supernova explosions.

In the preceding sections, we already emphasized the major role played
by stars in various aspects of the ISM, notably in its extremely
heterogeneous character, in its continual agitation, in its partial ionization,
in its heating, and in the formation and acceleration of \crs.
The present section taught us that stars are also key agents
in the generation and amplification of \is \mfs. Not only do they
constitute potential candidates for triggering the Galactic dynamo,
but also, and more importantly, they very likely drive much 
of the turbulence necessary for the alpha-effect and, hence, 
for dynamo action.

\section*{Acknowledgments}
I wish to thank both referees, C. Heiles and J. M. Shull,
for their careful reading of the paper and for their numerous suggestions, 
which greatly helped improve its content.
I am also grateful to T. M. Dame, J. M. Dickey, L. M. Haffner, 
M. H. Heyer, and B. P. Wakker,
who kindly provided me with the maps and spectra appearing 
in Figs.~\ref{CO_map}, \ref{HI_spectra}, \ref{HI_map}, \ref{Halpha_map}, 
and \ref{spectra}.

\section*{Appendix}

The purpose of this Appendix is to explain how the large-scale spatial
distribution of an \is gas component can be mapped by means of one of
its radio emission lines (say, the 21-cm emission line of H~{\sc i}),
once a model of Galactic rotation has been adopted (see Fig.~\ref{galaxy}).
Radio spectra including the chosen emission line are taken in a large
number of directions scanning the sky.
For each direction, the emission line spreads out over a range
of wavelengths around the rest wavelength, $\lambda_0$, as illustrated 
in Fig.~\ref{line}(a) (with $\lambda_0 = 21$~cm in the case of H~{\sc i}).

If the \is gas is transparent or, in more astrophysical terms, 
optically thin for the considered line (as often assumed for the \hy 
21-cm emission line), photons emitted toward the Earth travel 
all the way to the receiver without being re-absorbed by the gas,
so that they automatically contribute to the line intensity.
Therefore, the specific intensity, $I_\lambda$, at any given wavelength
$\lambda$ is directly proportional to the amount of material producing
the emission at that wavelength. The latter can be traced back to
a given location in the Galaxy, by noting that the Doppler shift,
$(\lambda - \lambda_0) / \lambda_0$, is straightforwardly related 
to the line-of-sight velocity with respect to the Earth, which, in turn, 
can be converted into a distance from us with the help of the Galactic 
rotation curve.
In fact, inside the solar circle, there exist two distances corresponding 
to a given line-of-sight velocity (for instance, 1 and $1^{'}$ 
in Fig.~\ref{galaxy} have the same line-of-sight velocity),
but this twofold distance ambiguity may, under some conditions,
be resolved with appropriate pattern recognition techniques.

If the considered line is optically thick (like the CO 2.6-mm emission line), 
most of the emitted photons are re-absorbed before reaching the receiver,
and the line intensity saturates at a level that is a function of
the gas temperature, independent of the amount of emitting material.
The case of the CO 2.6-mm emission line, however, is a little more subtle.
An idealized CO emission spectrum is sketched in Fig.~\ref{line}(b). 
Each feature in the spectrum can be attributed to a clump of emitting material 
in the observed direction, whose line-of-sight velocity is that deduced
from the Doppler shift of the feature. The fact that individual 
features are well separated indicates that the probability of overlap 
is small and that every clump along the line of sight is fully represented 
in the spectrum. Once averaged over a small area in the plane of the sky,
the CO emission spectrum has a profile similar to that drawn in
Fig.~\ref{line}(a),
where the contribution from every emitting clump is fully taken into account.
Thus, even though individual features are optically thick, the average
emission line is effectively optically thin, and its specific intensity
as a function of wavelength provides a direct measure of the amount of CO 
as a function of line-of-sight velocity, which, as before, translates
into a distance from us -- with a twofold ambiguity inside the solar circle.
For a more detailed discussion on the supposedly linear relationship
between the CO line intensity and its column density, the reader is
referred to the review paper by Scoville and Sanders (1987).

A few real emission spectra from recent Galactic surveys are displayed
in Fig.~\ref{spectra}.

\begin{table}
\caption{Descriptive parameters of the different components of the \is gas,
according to the references quoted in the main text.
$T$ is the temperature,
$n$ is the true (as opposed to space-averaged) number density of
hydrogen nuclei near the Sun, 
$\Sigma_\odot$ is the azimuthally-averaged mass density per unit area 
at the solar circle,
and $\cal M$ is the mass contained in the entire Milky Way.
Both $\Sigma_\odot$ and $\cal M$ include 70.4~\% of hydrogen, 
28.1~\% of helium, and 1.5~\% of heavier elements.
All values were rescaled to $R_\odot = 8.5$~kpc, in accordance with
footnote~\ref{note}.}
\begin{tabular}{lcccc} 
Component & $T$ (K) & $n$ (cm$^{-3}$) 
 & $\Sigma_\odot$ ($M_\odot$~pc$^{-2}$) & $\cal M$ ($10^9~M_\odot$) \\
\tableline
Molecular & $10 - 20 \ $ & $10^2 - 10^6$ & $\sim 2.5$ & 
 $\sim 1.3\tablenote{adapted from Bronfman \etalv 1988.} - 
       2.5\tablenote{adapted from Clemens \etalv 1988.}$ \\
Cold atomic & $50 - 100$ & $20 - 50$ & $\sim 3.5$ & \\
\noalign{\vspace{-6pt}}
& & & & {\large \}} $\gapr 6.0$ \qquad \qquad \quad \\
\noalign{\vspace{-6pt}}
Warm atomic & $6000 - 10000$ & $0.2 - 0.5$ & $\sim 3.5$ & \\
Warm ionized & $\sim 8000$ & $0.2 - 0.5$ & $\sim 1.4$ & 
 \phantom{\large \}} $\gapr 1.6$ \qquad \qquad \quad \\
Hot ionized & $\sim 10^6 \ $ & $\sim 0.0065$ & & \\
\end{tabular}
\label{t1}
\end{table}

\begin{figure}
\caption{Column density of \is hydrogen, defined as the number of hydrogen
nuclei contained in a vertical cylinder of unit cross section through
the Galactic disk, $N$, and mass density per unit area of \is matter
(including 70.4~\% of hydrogen, 28.1~\% of helium, and 1.5~\% of heavier
elements), $\Sigma = 1.42 \, m_{\rm P} \, N$, averaged over Galactocentric 
azimuthal angle, as a function of Galactic radius, $R$,
for the different gas components.
The solid lines give the contribution from the molecular gas
(thick line: from Bronfman \etalv 1988; thin line: from Clemens \etalv 1988),
the triple-dot--dashed line gives the contribution from the cold + warm atomic 
gas (from Lockman, 1984; Diplas and Savage, 1991; Dickey and Lockman, 1990),
and the dashed line gives the contribution from the ionized gas outside 
the traditional \h2 regions (from Cordes \etalv 1991, 
with a Gaussian radial scale length of 30~kpc for the thick component).}
\label{column_density}
\end{figure}

\begin{figure}
\caption{Space-averaged number density of \is hydrogen nuclei,
$\langle n \rangle$, and space-averaged mass density of \is matter
(including 70.4~\% of hydrogen, 28.1~\% of helium, and 1.5~\% of heavier
elements), $\langle \rho \rangle = 1.42 \, m_{\rm P} \, \langle n \rangle$,
as a function of Galactic height, $Z$, 
at the solar circle ($R = R_\odot$),
for the different gas components.
The solid lines give the contribution from the molecular gas
(thick line: from Bronfman \etalv 1988; thin line: from Clemens \etalv 1988),
the triple-dot--dashed line gives the contribution from the cold + warm atomic 
gas (from Dickey and Lockman, 1990),
and the dashed line gives the contribution from the ionized gas outside 
the traditional \h2 regions (from Cordes \etalv 1991, 
with a Gaussian radial scale length of 30~kpc for the thick component;
in agreement with Reynolds, 1991, with an exponential scale height of 1~kpc).}
\label{space_density}
\end{figure}

\begin{figure}
\caption{High-resolution CO map of a $10.6^\circ \times 8.4^\circ$ portion
of the sky centered on $(l = 112^\circ, \ b = 1^\circ)$,
from the {\it Five College Radio Astronomy Observatory (FCRAO)}
CO survey of the outer Galaxy (see Heyer \etalv 1998).
Figure courtesy of M. H. Heyer.}
\label{CO_map}
\end{figure}

\begin{figure}
\caption{Matched pair of \hy 21-cm emission (upper panel)
and absorption (lower panel) spectra, respectively near to and right in
the direction of the bright Galactic \h2 region G326.65+0.59,
from the {\it Southern Galactic Plane Survey} 
(see McClure-Griffiths \etalv 1999).
The $x$-axis is labeled in terms of line-of-sight velocity (km~s$^{-1}$),
after use has been made of the straightforward relationship between
Doppler shift, $(\lambda - \lambda_0) / \lambda_0$, and line-of-sight
velocity.
Figure courtesy of J. M. Dickey.}
\label{HI_spectra}
\end{figure}

\begin{figure}
\caption{High-resolution \hy maps of 
(a) a $120^\circ \times 30^\circ$ portion of the sky 
centered on $(l = 80^\circ, \ b = -40^\circ)$
at velocities between $-20$ and 20~km~s$^{-1}$,
from the {\it Leiden-Dwingeloo Survey} (Hartmann and Burton, 1997);
figure courtesy of B. P. Wakker;
(b) a $6^\circ \times 3^\circ$ portion of the sky 
centered on $(l = 260.5^\circ, \ b = 0^\circ)$
at velocity 50~km~s$^{-1}$,
from the {\it Southern Galactic Plane Survey}; 
figure courtesy of J. M. Dickey.}
\label{HI_map}
\end{figure}

\begin{figure}
\caption{High-resolution H$\alpha$ map of a $90^\circ \times 70^\circ$ 
portion of the sky centered on $(l = 115^\circ, \ b = 0^\circ)$
at velocities between $-60$ and 40~km~s$^{-1}$,
from the {\it WHAM} survey.
Figure courtesy of L. M. Haffner.}
\label{Halpha_map}
\end{figure}

\begin{figure}
\caption{Interstellar magnetic pressure, $P_{\rm M}$,
and \c_r pressure, $P_{\rm CR}$, as a function of Galactic height, $Z$,
at the solar circle ($R = R_\odot$), from Eqs.~(10) and (11), respectively.}
\label{pressure}
\end{figure}

\begin{figure}
\caption{Galactic supernova rate per unit area, $\sigma$,
as a function of Galactic radius, $R$, for both types of supernovae.
The Type~I supernova rate, $\sigma_{\rm I}$, follows from the stellar
disk model of Freeman (1987) and is given by Eq.~(14) (solid line).
For the Type~II supernova rate, $\sigma_{\rm II}$, we show both
an estimate based on the pulsar model of Johnston (1994)
(see Eq.~(16); dashed line) and an estimate based on the \h2 region
model of McKee and Williams (1997) (see main text; dotted line).}
\label{rate_area}
\end{figure}

\begin{figure}
\caption{Galactic supernova rate per unit volume, ${\cal R}$, 
as a function of Galactic height, $Z$, at the solar circle ($R = R_\odot$),
for both types of supernovae.
The Type~I supernova rate, ${\cal R}_{\rm I}$, follows from the stellar
disk model of Freeman (1987) and is given by Eq.~(15) (solid line).
The Type~II supernova rate, ${\cal R}_{\rm II}$, is based on 
the pulsar models of Johnston (1994) and Narayan and Ostriker (1990)
and is given by Eq.~(17) (dashed line).}
\label{rate_volume}
\end{figure}

\begin{figure}
\caption{Schematic view of our Galaxy seen from above.
GC indicates the position of the Galactic center 
and $\odot$ the position of the Sun.
$l$ is the Galactic longitude defined with respect to the Sun and
measured counterclockwise from the direction to the Galactic center.
I, II, III, and IV denote the first, second, third, and fourth 
Galactic quadrants.
The \is gas is assumed to rotate clockwise about the Galactic center
at a rate decreasing away from it.
In a frame rotating with the Sun, the gas interior to the solar orbit
rotates clockwise, while the exterior gas rotates counterclockwise,
as indicated by the thick arrows.}
\label{galaxy}
\end{figure}

\begin{figure}
\caption{Schematic profile of an emission line with rest wavelength,
$\lambda_0$, obtained in the direction $l$ drawn in Fig.~\ref{galaxy}.
$I_\lambda$ is the specific intensity at wavelength $\lambda$.
$\lambda_1$ corresponds to point 1 or $1^{'}$ in Fig.~\ref{galaxy},
while $\lambda_2$  corresponds to point 2.
(a) Generic case, directly applicable to the \hy 21-cm emission line
(adapted from Shu, 1982, p.~268).
(b) Case of the CO 2.6-mm emission line 
(adapted from Scoville and Sanders, 1987).}
\label{line}
\end{figure}

\begin{figure}
\caption{Real emission spectra, for which
the Doppler shift, $(\lambda - \lambda_0) / \lambda_0$, 
on the $x$-axis, has already been converted to line-of-sight velocity.
(a) Typical \hy emission spectra toward $(l = 90^\circ, \ b = 0^\circ)$
and $(l = 180^\circ, \ b = 10^\circ)$, 
from the {\it Leiden-Dwingeloo Survey} (courtesy of B. P. Wakker).
(b) Typical CO emission spectrum toward $(l = 20.7^\circ, \ b = 0^\circ)$,
obtained with the 1.2-m Telescope at the Harvard-Smithsonian Center 
for Astrophysics (courtesy of T. M. Dame).}
\label{spectra}
\end{figure}

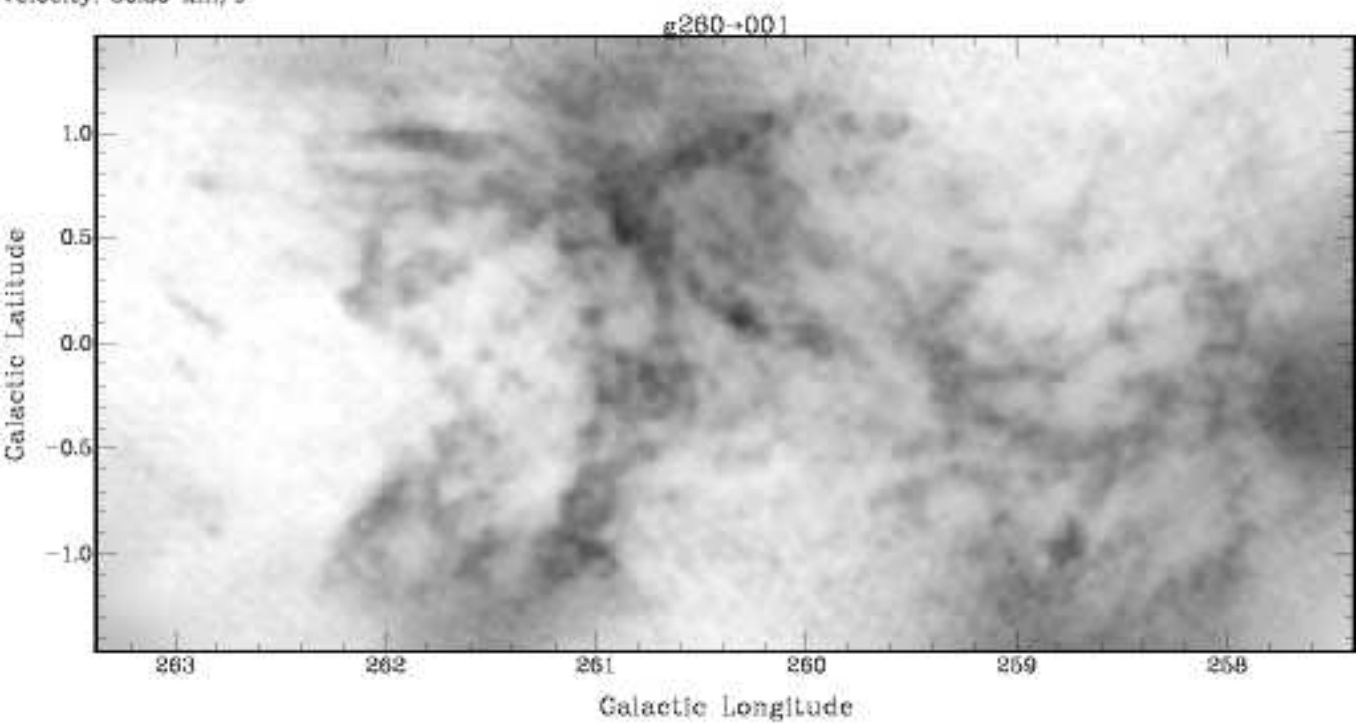
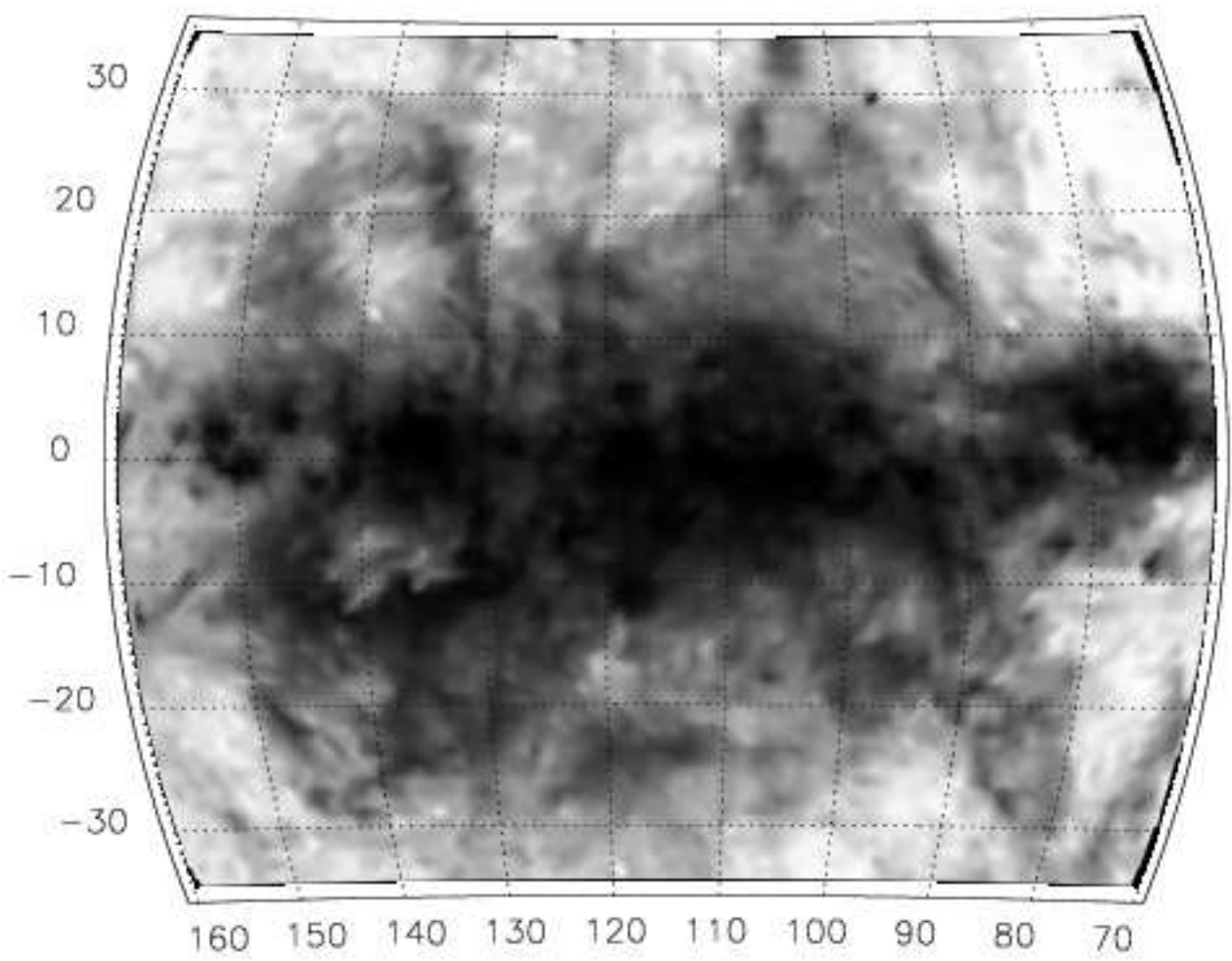
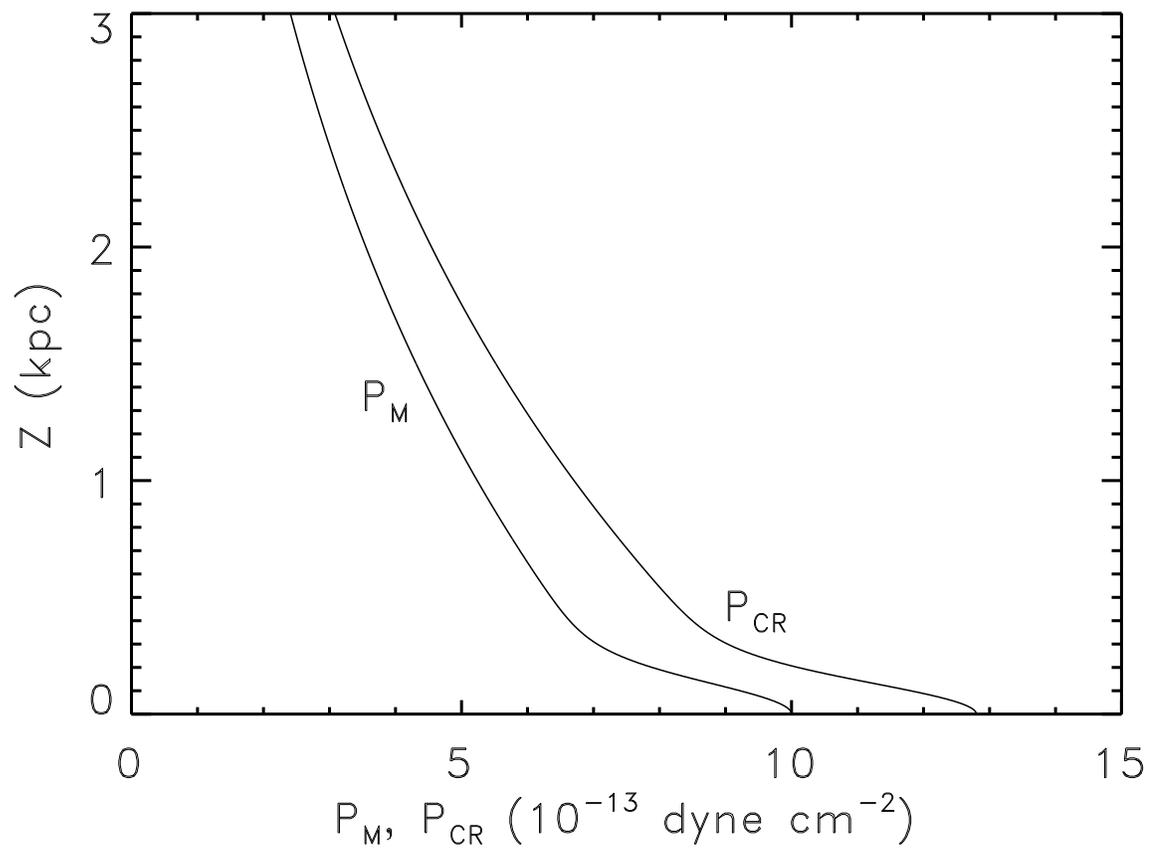
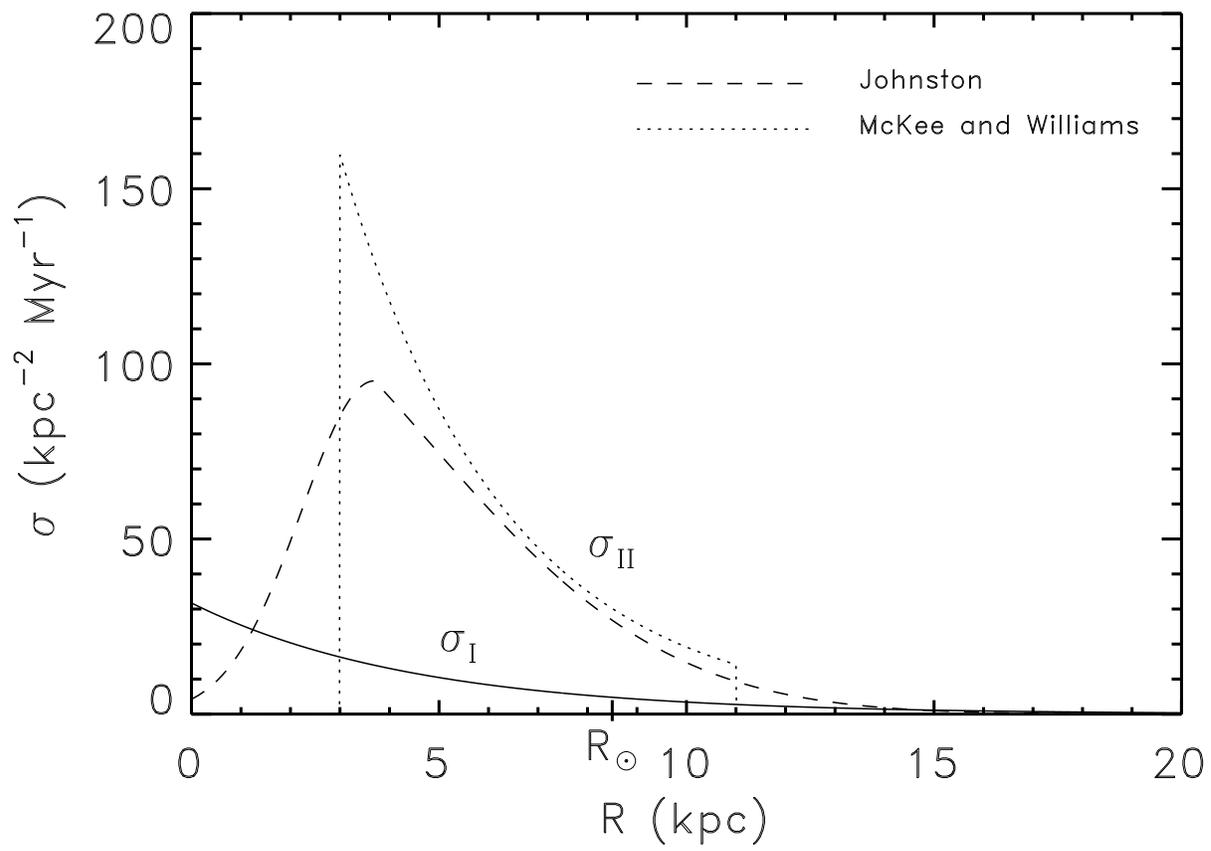
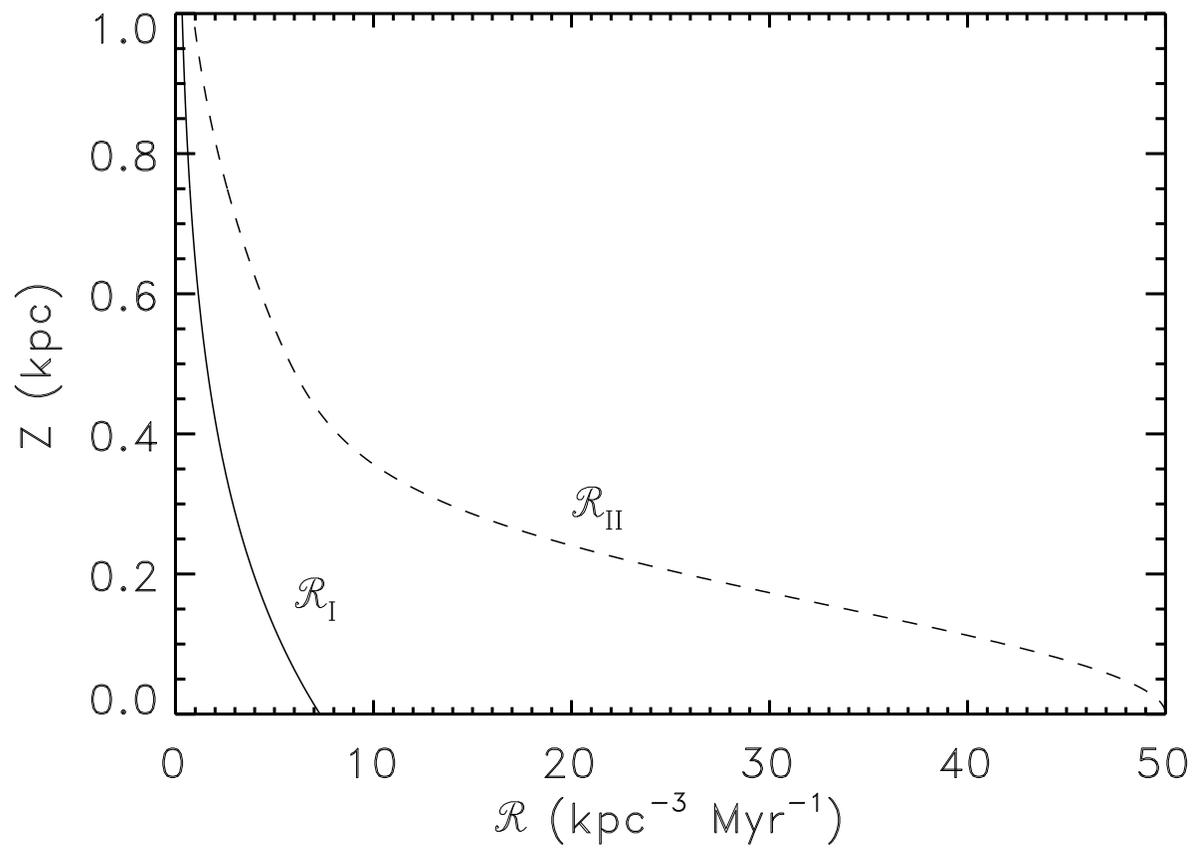
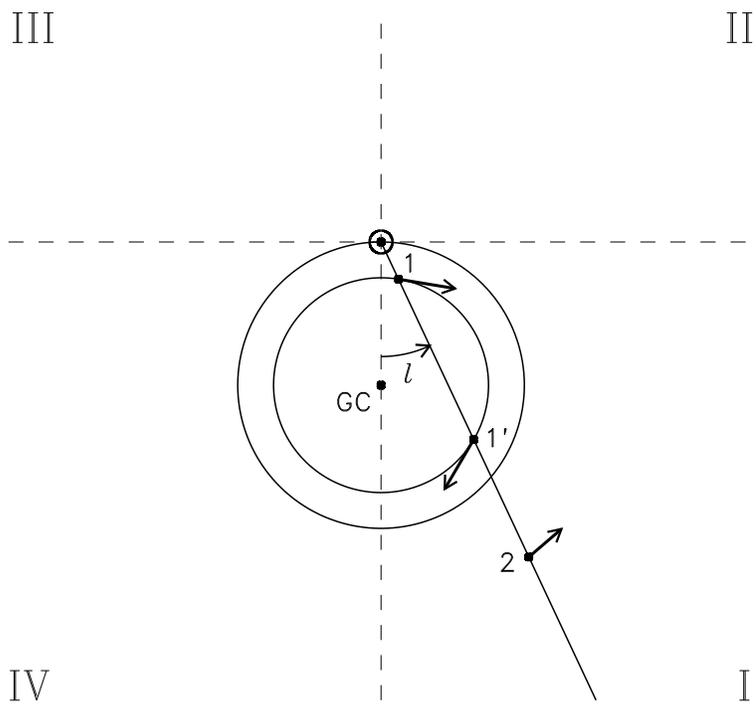
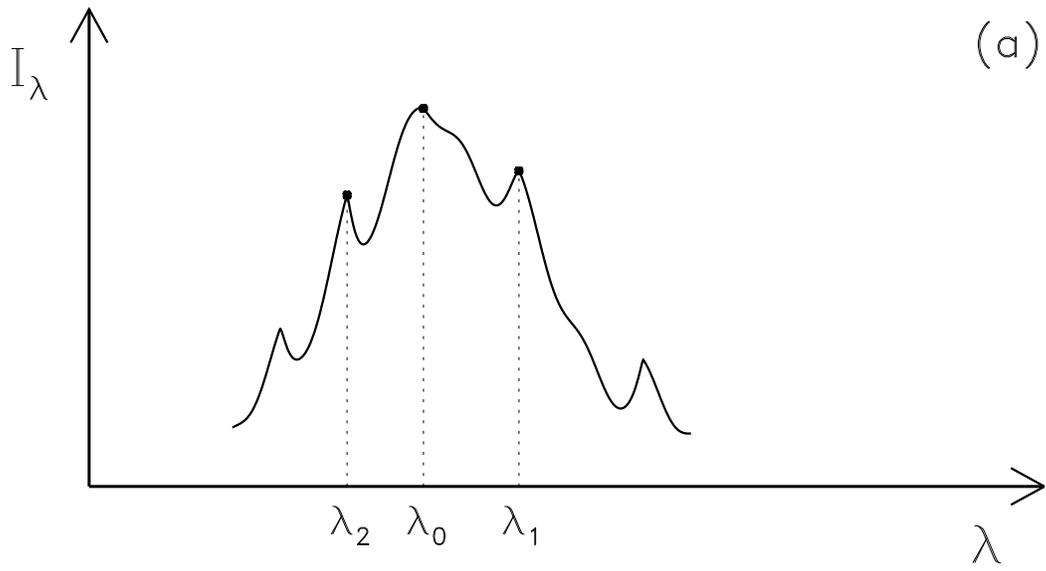
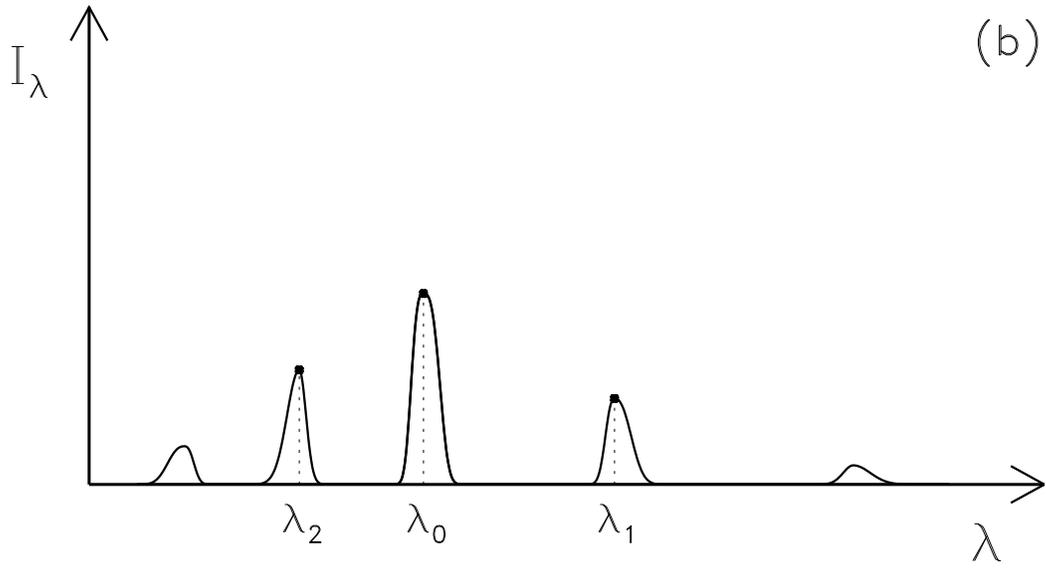
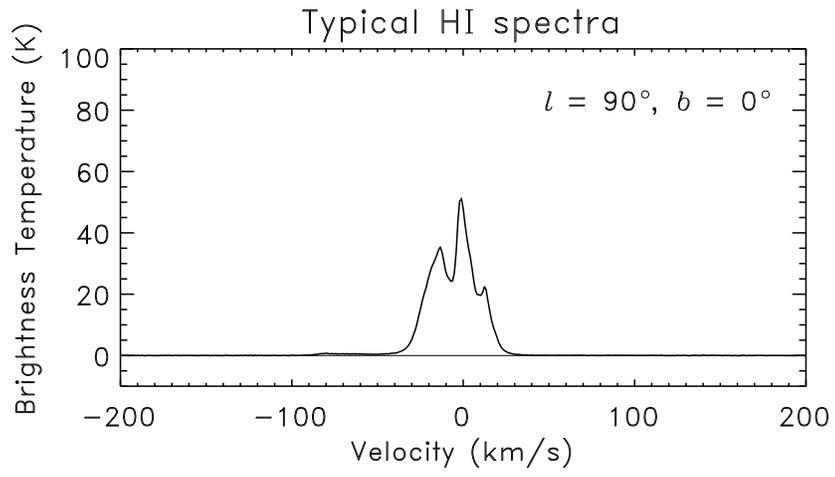
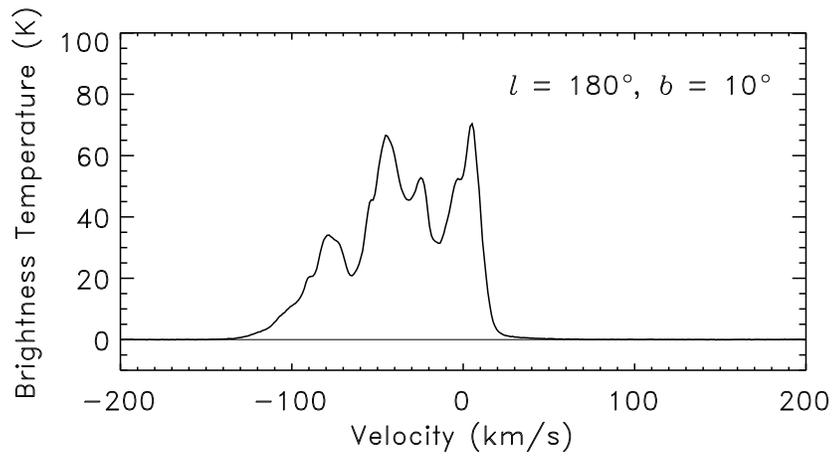
\end{document}